\documentclass[aps,prl,twocolumn,superscriptaddress]{revtex4-1}
\usepackage{amssymb, amsmath, bm}
\usepackage{graphicx,onlyamsmath}

\usepackage{natbib}
\usepackage{xcolor}
\usepackage[normalem]{ulem}  

\begin{document}

\title{Disorder-induced phase transition in Dirac systems beyond the linear approximation}

\author{Sergey S.~Krishtopenko}
\affiliation{Laboratoire Charles Coulomb (L2C), UMR 5221 CNRS-Universit\'{e} de Montpellier, F- 34095 Montpellier, France}

\author{Mauro Antezza}
\affiliation{Laboratoire Charles Coulomb (L2C), UMR 5221 CNRS-Universit\'{e} de Montpellier, F- 34095 Montpellier, France}
\affiliation{Institut Universitaire de France, 1 rue Descartes, F-75231 Paris Cedex 05, France}

\author{Fr\'{e}d\'{e}ric Teppe}
\email[]{frederic.teppe@umontpellier.fr}
\affiliation{Laboratoire Charles Coulomb (L2C), UMR 5221 CNRS-Universit\'{e} de Montpellier, F- 34095 Montpellier, France}
\date{\today}

\begin{abstract}
By using the self-consistent Born approximation, we investigate disorder effect induced by the short-range impurities on the band-gap in two-dimensional Dirac systems with the higher order terms in momentum. Starting from the Bernevig-Hughes-Zhang (BHZ) model, we calculate the density-of-states as a function of the disorder strength. We show that due to quadratic corrections to the Dirac Hamiltonian, the band-gap is always affected by the disorder even if the system is gapless in the clean limit. Finally, we explore the disorder effects by using an advanced effective Hamiltonian describing the side maxima of the valence subband in HgTe~quantum wells. We show that the band-gap and disorder-induced topological phase transition in the real structures may differ significantly from those predicted within the BHZ model.
\end{abstract}

\maketitle

\emph{Introduction.}--The rise of graphene~\cite{q1} has paved the way to the intensive investigation of Dirac fermions in condensed matter~\cite{q2}. Since then, Dirac fermion physics has also been analyzed in many other two-dimensional (2D) systems~\cite{q3}. From a general point of view, the presence of the massless Dirac cones is protected against any single-particle and many-body perturbations, at least as long as the interaction does not lead to a spontaneous breaking of symmetry~\cite{q19,q20,q21,q22,q23,q24}. The latter means that disorder cannot open the band-gap in the massless Dirac model with only the linear terms in momentum~\cite{q15a,q15b,q15c,q16,q17,q18}.

\begin{figure}
\includegraphics [width=0.95\columnwidth, keepaspectratio] {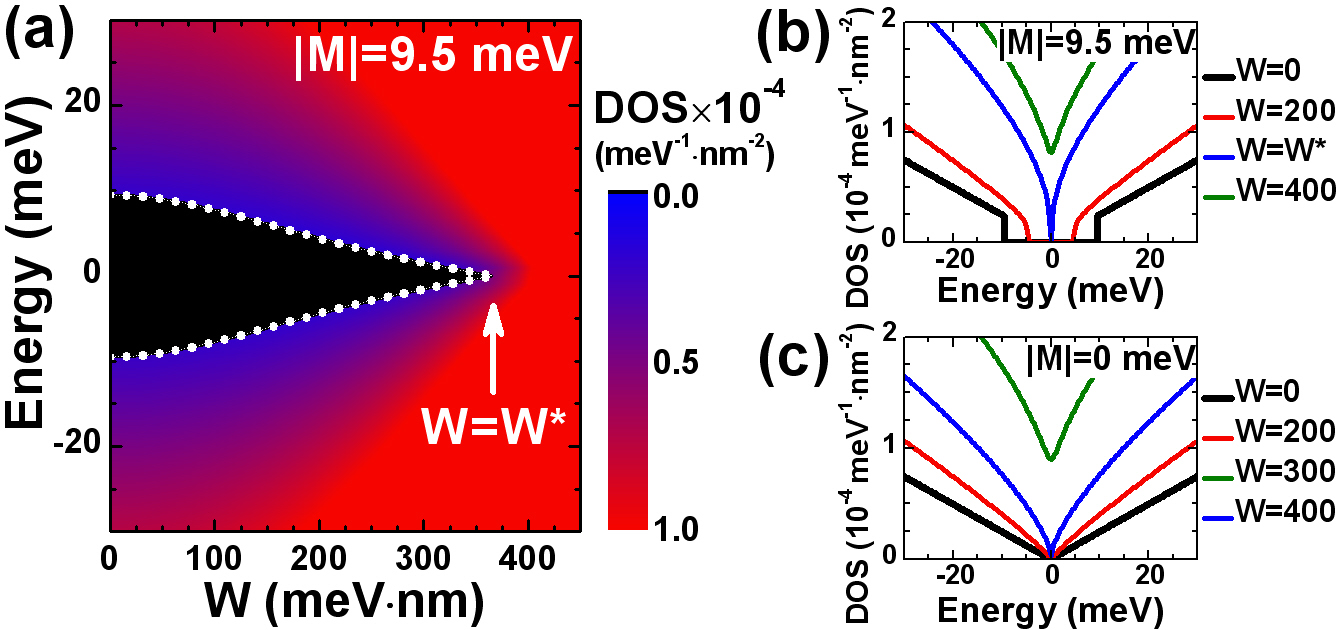} 
\caption{\label{Fig:1} (a) Color map of DOS as a function of the disorder strength~$W$ for the massive Dirac fermions with the energy $E=\pm\sqrt{M^2+A^2k^2}$, where $|M|=9.5$~meV and $A=358$~meV$\cdot$nm. The white curves represent the band edges with vanishing DOS described by Eq.~(\ref{eq:6}). (b,c) The DOS at different values of~$W$ for $|M|=9.5$~meV and $|M|=0$~meV.}
\end{figure}

Relatively less attention has been devoted to the study of disorder effects in the massive Dirac model~\cite{q25,q26,q27}. The analysis within the self-consistent Born approximation (SCBA) reveals a band-gap closing above a threshold of the disorder strength (see also Fig.~\ref{Fig:1}). Note that the strong disorder may also produce the onset of midgap impurity-induced states in highly disordered massive Dirac models~\cite{q28,q29,q30}, which needs a \emph{t}-matrix approach (beyond the SCBA scheme) to be revealed.

Many 2D systems, however, host Dirac fermions at small momentum only, while their description requires the terms beyond the linear approximation. Prominent examples are the surface states of three-dimensional topological insulators (3D~TIs)~\cite{q4,q5,q6} and their films~\cite{q31,q32,q33}. Another 2D systems are HgTe/CdHgTe~\cite{q9,q10,q11} and three-layer InAs/GaSb quantum wells (QWs)~\cite{q12,q13,q14}. All of them are described by the Bernevig-Hughes-Zhang (BHZ) Hamiltonian~\cite{q9}, in which the quadratic corrections to the Dirac model allow for the proper characterization of the topological states~\cite{q47}.

The role of disorder beyond the linear approximation became yet more complicated after numerical simulations of Li~\emph{et~al.}~\cite{q34}. By using a tight-binding version of the BHZ Hamiltonian, they have found that disorder may induce a novel phase with a quantized conductance called as \emph{topological Anderson insulator} (TAI)~\cite{q34}. Later, Groth~\emph{et~al.}~\cite{q35} have shown that formation of TAI is caused by the quadratic terms $\propto k^2\sigma_z$ in the BHZ Hamiltonian, which are absent for graphene even beyond the linear approximation~\cite{q2}. Moreover, it was shown that so contrary to the name "topological Anderson insulator", such weak-disorder topological transition is not an Anderson transition at all, and it can be treated within the SCBA~\cite{q35,q36,q37}. Although the mentioned works~\cite{q34,q35,q36,q37} are based on the tight-binding calculations on the square lattice with the constant $a$ (typically $a=5$~nm~\cite{q34,q35,q36,q37}), they indicate that the disorder effects in the BHZ Hamiltonian may differ significantly from those known in the linear Dirac model.

In this work, we investigate how disorder changes the band-gap in Dirac systems beyond the linear approximation. By using the SCBA, we directly calculate the density-of-states (DOS) within the \emph{continuous} BHZ Hamiltonian and more advanced model~\cite{q38} describing the side maxima of the valence band in HgTe~QWs. Our results univocally demonstrate a crucial role of the high-order terms in the disorder effects.

\emph{The two-band BHZ model and SCBA.}--The low-energy BHZ Hamiltonian has the form
\begin{equation}
\label{eq:1}
H_{\mathrm{2D}}(\mathbf{k})=\begin{pmatrix}
H_{\mathrm{BHZ}}(\mathbf{k}) & 0 \\ 0 & H_{\mathrm{BHZ}}^{*}(-\mathbf{k})\end{pmatrix},
\end{equation}
where asterisk stands for complex conjugation, $\mathbf{k}=(k_x,k_y)$ is the momentum in the plane, and $H_{\mathrm{BHZ}}(\mathbf{k})=\epsilon_{k}\mathbf{I}_2+d_a(\mathbf{k})\sigma_a$. Here, $\mathbf{I}_2$ is a 2$\times$2 unit matrix, $\sigma_a$ are the Pauli matrices, $\epsilon_{k}=C-Dk^2$, $d_1(\mathbf{k})=-Ak_x$, $d_2(\mathbf{k})=-Ak_y$, $d_3(k)=M-Bk^2$ and $k^2=k_x^2+k_y^2$. In the QWs case, the mass parameter $M$ describes inversion between the electron-like \emph{E}1 and hole-like \emph{H}1 subbands: $M>0$ corresponds to a trivial state, while $M<0$ for a quantum spin Hall insulator (QSHI) state~\cite{q9}. For the surface states of 3D~TIs, non-vanishing $M$ conforms to the gap opened due to the tunnel-coupling between the opposite surfaces in the thin films~\cite{q31,q32,q33}. A block-diagonal form of $H_{2D}(\mathbf{k})$ in Eq.~(\ref{eq:1}) (cf. Refs~\cite{q39,q40}) allows to focus on the upper block only, while the calculations for the lower block are performed in the same way.

In order to calculate DOS in the presence of disorder, we add the random impurity potential to $H_{\mathrm{BHZ}}(\mathbf{k})$:
\begin{eqnarray}
\label{eq:2}
V_{imp}(\textbf{r})=\sum_{j}v(\textbf{r}-\textbf{R}_j),~v(\textbf{r})=\int\dfrac{d^2\textbf{q}}{(2\pi)^2}\tilde{v}(\textbf{q})e^{i\textbf{q}\cdot\textbf{r}},~~~~~
\end{eqnarray}
where $R_j$ denotes position of impurities and $v(\textbf{r})$ is the potential of an individual impurity, which is assumed to be isotropic, i.e., $\tilde{v}(\textbf{q})=\tilde{v}(q)$ with $|\textbf{q}|=q$. Then, we start from the Dyson equation in the momentum representation for the disorder-averaged Green's function $\hat{G}(\mathbf{k},\varepsilon)$ and the self-energy matrix $\hat{\Sigma}(\mathbf{k},\varepsilon)$ considered in the SCBA, and illuminate the dependence on the direction of $\mathbf{k}$ by applying a unitary transformation such as $\tilde{H}_{\mathrm{BHZ}}(k)=U(\theta_{\mathbf{k}})H_{\mathrm{BHZ}}(\mathbf{k})U(\theta_{\mathbf{k}})^{-1}$. After some calculations provided in the Supplemental Materials~\cite{SM}, we get the following self-consistent equations:
\begin{eqnarray}
\label{eq:3}
\hat{\Sigma}(k,\varepsilon)=n_{i}\int\limits_0^{K_c}\dfrac{k^\prime dk^\prime}{2\pi}\begin{pmatrix}
V_0(k,k^\prime)^2G_{11}^\prime & V_{-1}(k,k^\prime)^2G_{12}^\prime \\ V_{+1}(k,k^\prime)^2G_{21}^\prime & V_0(k,k^\prime)^2G_{22}^\prime \end{pmatrix},\notag\\
V_n(k,k^\prime)^2=\int\limits_0^{2\pi}\dfrac{d\theta}{2\pi}|\tilde{v}(\textbf{k}-\textbf{k}^\prime)|^2
\cos n\theta,~~~~~~~~~~~
\end{eqnarray}
where $n_{i}$ is the concentration of impurities, and $G_{ij}^\prime\equiv G_{ij}(k^\prime,\varepsilon)$ are the component of the Green's function $\hat{G}(k,\varepsilon)=[\varepsilon-\tilde{H}_{\mathrm{BHZ}}(k)-\hat{\Sigma}(k,\varepsilon)]^{-1}$. In Eqs.~(\ref{eq:3}), we introduce a cut-off wave-vector $K_c=\pi/a_{\mathrm{0}}$ (where $a_{\mathrm{0}}$ is the lattice constant), corresponding to the size of the Brillouin zone (cf. Refs~\cite{q25,q26,q27}). Once the Green's function is known, the DOS can be calculated as:
\begin{eqnarray}
\label{eq:4}
D(\varepsilon)=-\dfrac{g_S}{\pi}\int\limits_0^{K_c}\dfrac{k dk}{2\pi}\textrm{Im}\left\{\textrm{Tr}\left(\hat{G}(k,\varepsilon+i0)\right)\right\},
\end{eqnarray}
where the factor $g_S=2$ takes into account the contribution from the lower block in Eq.~(\ref{eq:1}).

To proceed further, we assume $\tilde{v}(q)=u_0$,
which corresponds to the disorder formed by the short-range impurities~\cite{q25,q26,q27}. In this case, the self-energy matrix is independent of $k$ and has the form $\hat{\Sigma}(\varepsilon)=\Sigma_0(\varepsilon)\mathbf{I}_2+\Sigma_z(\varepsilon)\sigma_z$. Under these conditions, the set in Eq.~(\ref{eq:3}) is written as
\begin{equation}
\label{eq:5}
\Sigma_0=\dfrac{W^2}{4\pi}\int\limits_0^{K_c^2}
\dfrac{X+Dx}
{\Lambda(x,\varepsilon)}dx,~~
\Sigma_z=\dfrac{W^2}{4\pi}\int\limits_0^{K_c^2}
\dfrac{Y-Bx}
{\Lambda(x,\varepsilon)}dx,
\end{equation}
where $W$ is a disorder strength defined as $W^2=n_{i}u_0^2$, $X(\varepsilon)=\varepsilon-C-\Sigma_0(\varepsilon)$, $Y(\varepsilon)=M+\Sigma_z(\varepsilon)$ and $\Lambda(x,\varepsilon)=(D^2-B^2)x^2+(2BY+2DX-A^2)x+X^2-Y^2$. Note that the above integrals are calculated analytically~\cite{SM}, transforming Eq.~(\ref{eq:5}) into the set of algebraic equations numerically solved by simple iterations.

First, we consider the case of linear Dirac model, corresponding to zero values of $B$ and $D$. As shown in Fig.~\ref{Fig:1}, the band-gap of the massive Dirac fermions decreases by increasing $W$ until it vanishes above a critical value $W^{*}$. Such behavior was also investigated previously~\cite{q25,q26,q27}. For the gapless system, disorder does not open a band-gap as it was shown before for graphene~\cite{q15a,q15b,q15c,q16,q17,q18}. This is also seen from Eq.~(\ref{eq:5}), as $\Sigma_z(\varepsilon)=0$ is the self-consistent solution at $M=0$ and $B=D=0$. Note that the changes of DOS with $W$ is independent of the sign of $M$ in the linear model.

\begin{figure*}
\includegraphics [width=1.90\columnwidth, keepaspectratio] {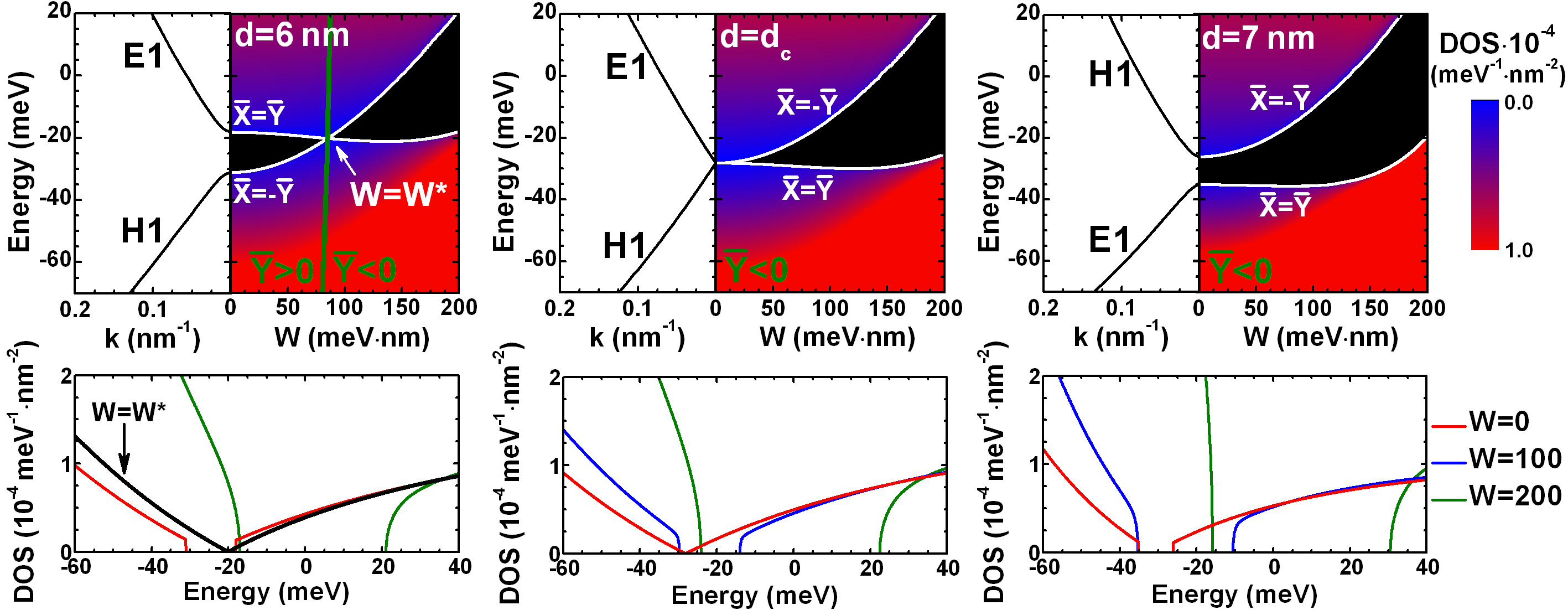} 
\caption{\label{Fig:2} Band structure and color map of the DOS as a function of the disorder strength~$W$ calculated in the two-band BHZ model for HgTe QW at different QW width: $d=6$~nm ($M>0$), $d=d_c$ ($M=0$) and $d=7$~nm ($M<0$). The band parameters are provided in the Supplemental Materials~\cite{SM}. The white curves represent the edges of the area with the vanishing DOS described by Eq.~(\ref{eq:6}). The green curve is found from $\overline{Y}(\varepsilon)=0$. The bottom panels show the DOS at several values of~$W$.}
\end{figure*}

The situation changes dramatically if we consider the square terms $\propto Bk^2\sigma_z$. Although we also include the terms $\propto Dk^2\mathbf{I}_2$ representing the electron-hole asymmetry, the DOS evolution remains qualitatively the same even as for $D=0$. Further, we focus on HgTe QWs, which require non-vanishing $D$ for their description~\cite{q38,q41}. As the band ordering in HgTe QWs is affected by hydrostatic pressure~\cite{q42}, temperature~\cite{q43,q44} and strain~\cite{q45,q46}, we note to consider HgTe/Cd$_{0.7}$Hg$_{0.3}$Te QWs grown on (001) CdTe buffer at zero temperature and pressure~\cite{q42}.

Figure~\ref{Fig:2} shows the evolution of DOS with the disorder strength $W$ for the HgTe QWs of different QW width. As it is seen, in contrast to the linear model, now the band-gap evolution strongly depends of the sign of $M$. If $M>0$, the band-gap decreases with $W$ and vanishes at a critical value $W^{*}$ and, than, it is re-opened again at $W>W^{*}$. Such behavior represents the disorder-induced topological phase transition previously discovered in the tight-binding calculations~\cite{q34,q35}. Let us now analyze it within the \emph{continuous} BHZ model.

Since the finite DOS is associated with a finite imaginary part of the functions $\Sigma_0(\varepsilon)$ and $\Sigma_z(\varepsilon)$, the band-gap region is characterized by the solution of Eq.~(\ref{eq:5}) with purely real quantities $\overline{\Sigma}_0(\varepsilon)$ and $\overline{\Sigma}_z(\varepsilon)$. The band edges can be obtained by solving the following equations:
\begin{eqnarray}
\label{eq:6}
\overline{X}(\varepsilon)=\overline{Y}(\varepsilon),~~\overline{X}(\varepsilon)=-\overline{Y}(\varepsilon),~~
\end{eqnarray}
where the upper bar stresses the values found on the set of \emph{real} numbers.
As seen from Fig.~\ref{Fig:2}, two curves described by Eq.~(\ref{eq:6}) cross at the transition point $W=W^{*}$, where $\overline{Y}(\varepsilon)$ changes the sign. As shown by Groth~\emph{et~al.}~\cite{q35}, $\overline{Y}(\varepsilon)$ has a meaning of the renormalized topological mass and its negative sign corresponds to the TAI state. We note that the disorder-induced phase transition at $M>0$ in Fig.~\ref{Fig:2} is caused by the negative values of $B$ in HgTe QWs~\cite{q9,q40,q41}, resulting to $\overline{\Sigma}_z(\varepsilon)<0$. In the systems with $B>0$, such transition arises at $M<0$.

As mentioned above, the disorder does not open the gap for the \emph{linear} massless Dirac fermions. Fig.~\ref{Fig:2} demonstrates that due the square terms $\propto Bk^2\sigma_z$, the gapless state becomes a critical state with $W^{*}=0$, and the band-gap is now affected by the disorder. Interestly, one may conclude that since the surface states of 3D~TIs are described by the BHZ Hamiltonian, they are not robust to the surface disorder. However, the parameters $M$ and $B$ are not independent for 3D~TIs~\cite{q31,q32,q33}. In the absence of the tunnel-coupling between the opposite surfaces, $M=0$ but $B$ vanishes as well~\cite{q31,q32}. The latter prevents the band-gap opening by the disorder.

We have considered a role of the square terms in the disorder-induced topological phase transition in Dirac systems. Further, we investigate how the higher-order terms beyond the BHZ model affect the band-gap in the real structures. These terms are crucial for the side maxima (SM) of the valence subband in HgTe QWs~\cite{q38,q42}

\begin{figure*}
\includegraphics [width=1.90\columnwidth, keepaspectratio] {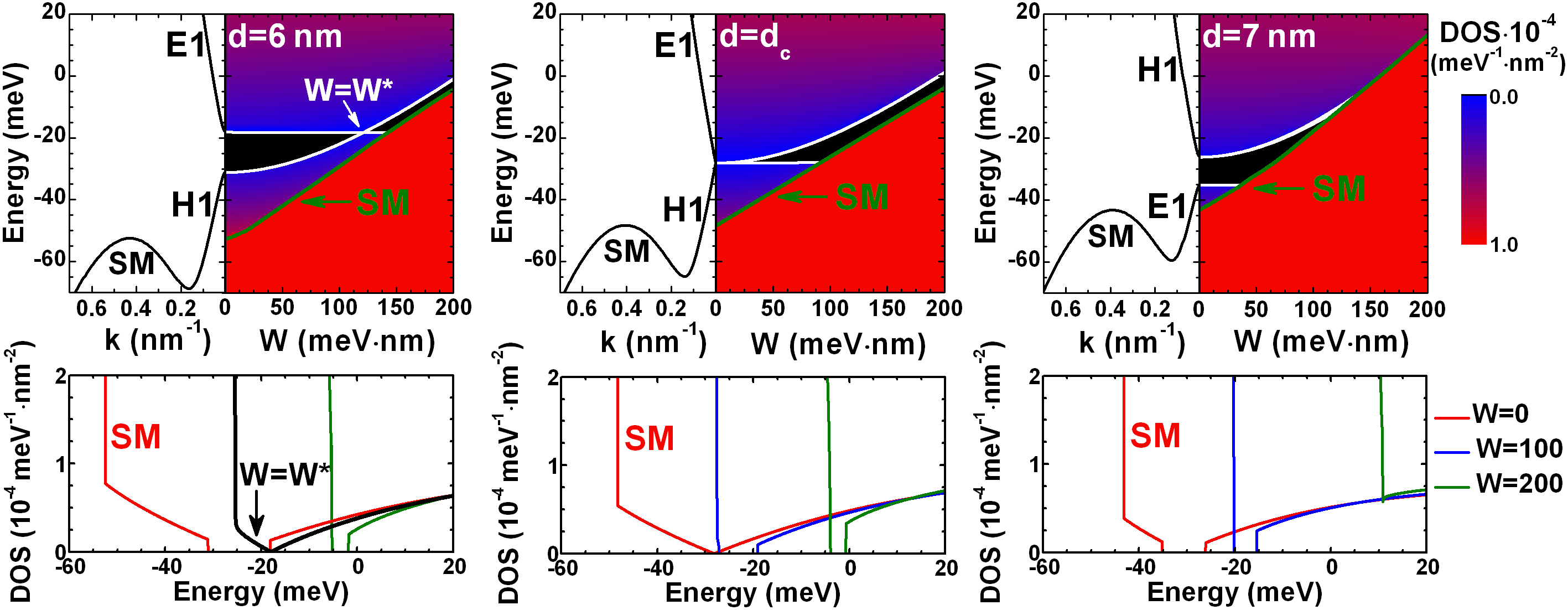} 
\caption{\label{Fig:3} Band structure and the DOS as a function of the disorder strength~$W$ calculated in the four-band 2D model~\cite{q38} for HgTe QW at different QW width: $d=6$~nm, $d=d_c$ and $d=7$~nm. The band parameters are provided in the Supplemental Materials~\cite{SM}. The white curves represent the edges of the area with the vanishing DOS found from the numerical calculations. The green curve shows the energy edge of the side maxima identified in the DOS. The bottom panels represent the DOS at several values of~$W$ for each of the QW width.}
\end{figure*}

\emph{The four-band 2D model and SCBA.}--The advanced Hamiltonian for HgTe QWs including the second electron-like \emph{E}2 and hole-like \emph{H}2 subbands is written as:
\begin{equation}
\label{eq:7}
H_{2D}(\mathbf{k})=\begin{pmatrix}
H_{4\times4}(\mathbf{k}) & 0 \\ 0 & H_{4\times4}^{*}(-\mathbf{k})\end{pmatrix}
\end{equation}
with the blocks $H_{4\times4}(\mathbf{k})$ and $H_{4\times4}^{*}(-\mathbf{k})$ defined as
\begin{equation}
\label{eq:8}
H_{4\times4}(\mathbf{k})=\begin{pmatrix}
\epsilon_{\mathbf{k}}+d_3(\mathbf{k}) & -Ak_{+} & R_{1}k_{-}^2 & S_{0}k_{-}\\
-Ak_{-} & \epsilon_{\mathbf{k}}-d_3(\mathbf{k}) & 0 & R_{2}k_{-}^2\\
R_{1}k_{+}^2 & 0 & \epsilon_{H2}(\mathbf{k})  & A_{2}k_{+}\\
S_{0}k_{+} & R_{2}k_{+}^2 & A_{2}k_{-} & \epsilon_{E2}(\mathbf{k}) \end{pmatrix},
\end{equation}
where $\epsilon_{E2}(\mathbf{k})=C+M+\Delta_{E1E2}+B_{E2}(k_x^2+k_y^2)$, $\epsilon_{H2}(\mathbf{k})=C-M-\Delta_{H1H2}+B_{H2}(k_x^2+k_y^2)$, $\Delta_{E1E2}$ and $\Delta_{H1H2}$ are the gaps between the \emph{E}1 and \emph{E}2 subbands and the \emph{H}1 and \emph{H}2 subbands, respectively~\cite{q38}.

Then, with a unitary transformation such as $\tilde{H}_{4\times4}(k)=V(\theta_{\mathbf{k}})H_{4\times4}(\mathbf{k})V(\theta_{\mathbf{k}})^{-1}$, the self-energy matrix $\hat{\Sigma}_{4\times4}(k,\varepsilon)$ in the SCBA has the form~\cite{SM}:
\begin{eqnarray}
\label{eq:9}
\hat{\Sigma}_{4\times4}(k,\varepsilon)=n_{i}\int\limits_0^{K_c}\dfrac{k^\prime dk^\prime}{2\pi}~~~~~~~~~~~~~~~~~~~~~~~~~~~~~~~~~~~\notag\\
\times\begin{pmatrix}
V_0^2G_{11}^\prime & V_{-1}^2G_{12}^\prime & V_{+2}^2G_{13}^\prime & V_{+1}^2G_{14}^\prime \\[2pt]
V_{+1}^2G_{21}^\prime & V_{0}^2G_{22}^\prime & V_{+3}^2G_{23}^\prime & V_{+2}^2G_{24}^\prime \\[2pt]
V_{-2}^2G_{31}^\prime & V_{-3}^2G_{32}^\prime & V_{0}^2G_{33}^\prime & V_{-1}^2G_{34}^\prime \\[2pt]
V_{-1}^2G_{41}^\prime & V_{-2}^2G_{42}^\prime & V_{+1}^2G_{43}^\prime & V_{0}^2G_{44}^\prime
\end{pmatrix},~
\end{eqnarray}
where $n_i$, $K_c$ and $V_n(k,k^\prime)^2$ are the same as those for Eq.~(\ref{eq:3}), while $G_{ij}^\prime\equiv G_{ij}(k^\prime,\varepsilon)$ are the component of the averaged Green's function $\hat{G}(k,\varepsilon)=[\varepsilon-\tilde{H}_{4\times4}(k)-\hat{\Sigma}(k,\varepsilon)]^{-1}$. In the case of the short-range impurities, the self-energy matrix is diagonal and independent of $\varepsilon$ and Eq.~(\ref{eq:9}) transforms into the set of algebraic equations numerically solved by iteration procedure~\cite{SM}.

Figure~\ref{Fig:3} shows the DOS evolution with the disorder for the same QW widths as in Fig.~\ref{Fig:2}. As it is seen for the 6~nm QW, the disorder-induced phase transition at $W=W^{*}$ is still identified. However, the values of $W^{*}$ and the areas with the vanishing DOS differ significantly in two models. Particularly, the renormalized band-gap in the BHZ model may even exceed the largest gap known for the HgTe QWs~\cite{q45}, while the four-band model predicts the lower values.

Another feature, which can not be addressed in the BHZ model, is the evolution of the DOS associated with the side maxima (SM) of the top valence subband. In the clean limit, the side maxima result in the large step-like increasing of the DOS. At non-zero $W$, such step-like behavior can be also used for qualitative determination of the SM position. The white and green curves in the top panels of Fig.~\ref{Fig:3} represent the evolution of the band edges in the $\Gamma$ point and the side maxima, respectively. In contrast to the BHZ model, these curves can be identified only in the numerical calculations. As the SM position primarily depends on the distance between \emph{E}2 and \emph{H}2 subbands~\cite{q38}, its evolution with the disorder remains qualitatively the same for any values of $M$.

\begin{figure}
\includegraphics [width=0.90\columnwidth, keepaspectratio] {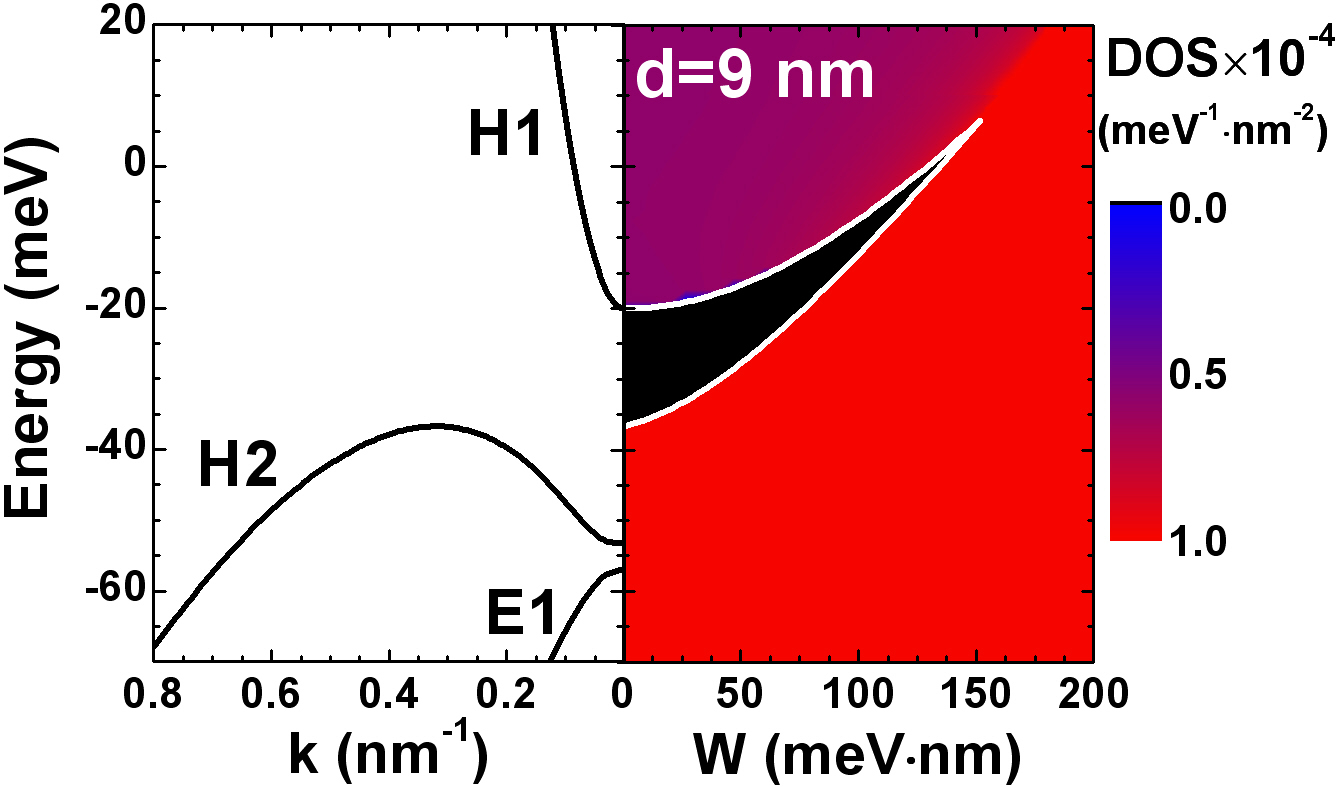} 
\caption{\label{Fig:4} Band structure and the DOS as a function of the disorder strength~$W$ for the 9~nm wide HgTe QW. The white curves represent the edges of the black area with the vanishing DOS identified in the numerical calculations.}
\end{figure}

As it is seen, the SM contribution increases with the disorder and strongly affects the area with the vanishing DOS. For the inverted HgTe QWs, the SM may result in the band-gap closing and transition into the semimetal state. The latter is clearly seen for the 9~nm wide HgTe QW representing \emph{indirect-gap} QSHI (see Fig~\ref{Fig:4}). Indeed, the upper boundary of the area with the vanishing DOS represents the evolution of the conduction band edge, while the lower boundary corresponds to the SM evolution. The semimetal state arises when the side maxima exceed the conduction band bottom. Note that such state also exists in the wide HgTe QWs in the clean limit~\cite{q48,q49}. Thus, the disorder may not only yield to the band-inversion as explained by Groth~\emph{et~al.}~\cite{q35} but induce the semimetal state as well.

To clarify if the disorder range in Figs~\ref{Fig:2}--\ref{Fig:4} is relevant for the HgTe QWs, we calculate the electron mobility $\mu_{W}$ caused by the short-range impurities~\cite{SM}. The calculations, performed in the relaxation time approximation~\cite{q51} within the BHZ model, evidence that $W<200$ corresponds to $\mu_{W}>4$~m$^2$/V$\cdot$s for the electron concentration $n_S=10^{11}$~cm$^{-2}$. This is comparable with the experimental values for HgTe QWs~\cite{q50}. Note that the mobility evaluation in the BHZ model is quit reliable for the conduction band, since it requires the description of electronic states only in the vicinity of the Fermi level, in contrast to the self-energy matrix, which is calculated over the whole Brillouin zone.

\emph{Conclusions.}--We have investigated the disorder effect caused by the short-range impurities on the band-gap and DOS in Dirac systems beyond the linear approximation. By using the SCBA and \emph{continuous} BHZ model, we show that the quadratic corrections to the Dirac Hamiltonian always result in the band-gap renormalization even if the system is gapless in the clean limit. We have also explored the role of the high-order terms beyond the BHZ model~\cite{q35} in the disorder effect in HgTe QWs. We have shown that the disorder-induced phase transition in the real structures may differ significantly from those predicted within the BHZ model. Our findings thus clearly demonstrate the invalidity of the BHZ model for quantitative description of the disorder effects in HgTe QWs.


\begin{acknowledgments}
The authors acknowledge T. Dietl (Institute of Physics PAS, Warsaw) for his critical comments and discussions. This work was supported by MIPS department of Montpellier University through the "Occitanie Terahertz Platform", by the Languedoc-Roussillon region via the "Gepeto Terahertz platform" and by the French Agence Nationale pour la Recherche (Colector project).
\end{acknowledgments}


\begin{thebibliography}{53}%
\makeatletter
\providecommand \@ifxundefined [1]{%
 \@ifx{#1\undefined}
}%
\providecommand \@ifnum [1]{%
 \ifnum #1\expandafter \@firstoftwo
 \else \expandafter \@secondoftwo
 \fi
}%
\providecommand \@ifx [1]{%
 \ifx #1\expandafter \@firstoftwo
 \else \expandafter \@secondoftwo
 \fi
}%
\providecommand \natexlab [1]{#1}%
\providecommand \enquote  [1]{``#1''}%
\providecommand \bibnamefont  [1]{#1}%
\providecommand \bibfnamefont [1]{#1}%
\providecommand \citenamefont [1]{#1}%
\providecommand \href@noop [0]{\@secondoftwo}%
\providecommand \href [0]{\begingroup \@sanitize@url \@href}%
\providecommand \@href[1]{\@@startlink{#1}\@@href}%
\providecommand \@@href[1]{\endgroup#1\@@endlink}%
\providecommand \@sanitize@url [0]{\catcode `\\12\catcode `\$12\catcode
  `\&12\catcode `\#12\catcode `\^12\catcode `\_12\catcode `\%12\relax}%
\providecommand \@@startlink[1]{}%
\providecommand \@@endlink[0]{}%
\providecommand \url  [0]{\begingroup\@sanitize@url \@url }%
\providecommand \@url [1]{\endgroup\@href {#1}{\urlprefix }}%
\providecommand \urlprefix  [0]{URL }%
\providecommand \Eprint [0]{\href }%
\providecommand \doibase [0]{http://dx.doi.org/}%
\providecommand \selectlanguage [0]{\@gobble}%
\providecommand \bibinfo  [0]{\@secondoftwo}%
\providecommand \bibfield  [0]{\@secondoftwo}%
\providecommand \translation [1]{[#1]}%
\providecommand \BibitemOpen [0]{}%
\providecommand \bibitemStop [0]{}%
\providecommand \bibitemNoStop [0]{.\EOS\space}%
\providecommand \EOS [0]{\spacefactor3000\relax}%
\providecommand \BibitemShut  [1]{\csname bibitem#1\endcsname}%
\let\auto@bib@innerbib\@empty
\bibitem [{\citenamefont {Geim}\ and\ \citenamefont {Novoselov}(2007)}]{q1}%
  \BibitemOpen
  \bibfield  {author} {\bibinfo {author} {\bibfnamefont {A.}~\bibnamefont
  {Geim}}\ and\ \bibinfo {author} {\bibfnamefont {K.}~\bibnamefont
  {Novoselov}},\ }\href {\doibase 10.1038/nmat1849} {\bibfield  {journal}
  {\bibinfo  {journal} {Nature Mater.}\ }\textbf {\bibinfo {volume} {6}},\
  \bibinfo {pages} {183} (\bibinfo {year} {2007})}\BibitemShut {NoStop}%
\bibitem [{\citenamefont {Castro~Neto}\ \emph {et~al.}(2009)\citenamefont
  {Castro~Neto}, \citenamefont {Guinea}, \citenamefont {Peres}, \citenamefont
  {Novoselov},\ and\ \citenamefont {Geim}}]{q2}%
  \BibitemOpen
  \bibfield  {author} {\bibinfo {author} {\bibfnamefont {A.~H.}\ \bibnamefont
  {Castro~Neto}}, \bibinfo {author} {\bibfnamefont {F.}~\bibnamefont {Guinea}},
  \bibinfo {author} {\bibfnamefont {N.~M.~R.}\ \bibnamefont {Peres}}, \bibinfo
  {author} {\bibfnamefont {K.~S.}\ \bibnamefont {Novoselov}}, \ and\ \bibinfo
  {author} {\bibfnamefont {A.~K.}\ \bibnamefont {Geim}},\ }\href {\doibase
  10.1103/RevModPhys.81.109} {\bibfield  {journal} {\bibinfo  {journal} {Rev.
  Mod. Phys.}\ }\textbf {\bibinfo {volume} {81}},\ \bibinfo {pages} {109}
  (\bibinfo {year} {2009})}\BibitemShut {NoStop}%
\bibitem [{\citenamefont {Wehling}\ \emph {et~al.}(2014)\citenamefont
  {Wehling}, \citenamefont {Black-Schaffer},\ and\ \citenamefont
  {Balatsky}}]{q3}%
  \BibitemOpen
  \bibfield  {author} {\bibinfo {author} {\bibfnamefont {T.~O.}\ \bibnamefont
  {Wehling}}, \bibinfo {author} {\bibfnamefont {A.~M.}\ \bibnamefont
  {Black-Schaffer}}, \ and\ \bibinfo {author} {\bibfnamefont {A.~V.}\
  \bibnamefont {Balatsky}},\ }\href {\doibase 10.1080/00018732.2014.927109}
  {\bibfield  {journal} {\bibinfo  {journal} {Adv. Phys.}\ }\textbf {\bibinfo
  {volume} {63}},\ \bibinfo {pages} {1} (\bibinfo {year} {2014})}\BibitemShut
  {NoStop}%
\bibitem [{\citenamefont {Appelquist}\ \emph {et~al.}(1988)\citenamefont
  {Appelquist}, \citenamefont {Nash},\ and\ \citenamefont
  {Wijewardhana}}]{q19}%
  \BibitemOpen
  \bibfield  {author} {\bibinfo {author} {\bibfnamefont {T.}~\bibnamefont
  {Appelquist}}, \bibinfo {author} {\bibfnamefont {D.}~\bibnamefont {Nash}}, \
  and\ \bibinfo {author} {\bibfnamefont {L.~C.~R.}\ \bibnamefont
  {Wijewardhana}},\ }\href {\doibase 10.1103/PhysRevLett.60.2575} {\bibfield
  {journal} {\bibinfo  {journal} {Phys. Rev. Lett.}\ }\textbf {\bibinfo
  {volume} {60}},\ \bibinfo {pages} {2575} (\bibinfo {year}
  {1988})}\BibitemShut {NoStop}%
\bibitem [{\citenamefont {Khveshchenko}(2001)}]{q20}%
  \BibitemOpen
  \bibfield  {author} {\bibinfo {author} {\bibfnamefont {D.~V.}\ \bibnamefont
  {Khveshchenko}},\ }\href {\doibase 10.1103/PhysRevLett.87.246802} {\bibfield
  {journal} {\bibinfo  {journal} {Phys. Rev. Lett.}\ }\textbf {\bibinfo
  {volume} {87}},\ \bibinfo {pages} {246802} (\bibinfo {year}
  {2001})}\BibitemShut {NoStop}%
\bibitem [{\citenamefont {Vafek}\ and\ \citenamefont {Case}(2008)}]{q21}%
  \BibitemOpen
  \bibfield  {author} {\bibinfo {author} {\bibfnamefont {O.}~\bibnamefont
  {Vafek}}\ and\ \bibinfo {author} {\bibfnamefont {M.~J.}\ \bibnamefont
  {Case}},\ }\href {\doibase 10.1103/PhysRevB.77.033410} {\bibfield  {journal}
  {\bibinfo  {journal} {Phys. Rev. B}\ }\textbf {\bibinfo {volume} {77}},\
  \bibinfo {pages} {033410} (\bibinfo {year} {2008})}\BibitemShut {NoStop}%
\bibitem [{\citenamefont {Wang}\ \emph {et~al.}(2010)\citenamefont {Wang},
  \citenamefont {Fertig},\ and\ \citenamefont {Murthy}}]{q22}%
  \BibitemOpen
  \bibfield  {author} {\bibinfo {author} {\bibfnamefont {J.}~\bibnamefont
  {Wang}}, \bibinfo {author} {\bibfnamefont {H.~A.}\ \bibnamefont {Fertig}}, \
  and\ \bibinfo {author} {\bibfnamefont {G.}~\bibnamefont {Murthy}},\ }\href
  {\doibase 10.1103/PhysRevLett.104.186401} {\bibfield  {journal} {\bibinfo
  {journal} {Phys. Rev. Lett.}\ }\textbf {\bibinfo {volume} {104}},\ \bibinfo
  {pages} {186401} (\bibinfo {year} {2010})}\BibitemShut {NoStop}%
\bibitem [{\citenamefont {Kotov}\ \emph {et~al.}(2012)\citenamefont {Kotov},
  \citenamefont {Uchoa}, \citenamefont {Pereira}, \citenamefont {Guinea},\ and\
  \citenamefont {Castro~Neto}}]{q23}%
  \BibitemOpen
  \bibfield  {author} {\bibinfo {author} {\bibfnamefont {V.~N.}\ \bibnamefont
  {Kotov}}, \bibinfo {author} {\bibfnamefont {B.}~\bibnamefont {Uchoa}},
  \bibinfo {author} {\bibfnamefont {V.~M.}\ \bibnamefont {Pereira}}, \bibinfo
  {author} {\bibfnamefont {F.}~\bibnamefont {Guinea}}, \ and\ \bibinfo {author}
  {\bibfnamefont {A.~H.}\ \bibnamefont {Castro~Neto}},\ }\href {\doibase
  10.1103/RevModPhys.84.1067} {\bibfield  {journal} {\bibinfo  {journal} {Rev.
  Mod. Phys.}\ }\textbf {\bibinfo {volume} {84}},\ \bibinfo {pages} {1067}
  (\bibinfo {year} {2012})}\BibitemShut {NoStop}%
\bibitem [{\citenamefont {Popovici}\ \emph {et~al.}(2013)\citenamefont
  {Popovici}, \citenamefont {Fischer},\ and\ \citenamefont {von Smekal}}]{q24}%
  \BibitemOpen
  \bibfield  {author} {\bibinfo {author} {\bibfnamefont {C.}~\bibnamefont
  {Popovici}}, \bibinfo {author} {\bibfnamefont {C.~S.}\ \bibnamefont
  {Fischer}}, \ and\ \bibinfo {author} {\bibfnamefont {L.}~\bibnamefont {von
  Smekal}},\ }\href {\doibase 10.1103/PhysRevB.88.205429} {\bibfield  {journal}
  {\bibinfo  {journal} {Phys. Rev. B}\ }\textbf {\bibinfo {volume} {88}},\
  \bibinfo {pages} {205429} (\bibinfo {year} {2013})}\BibitemShut {NoStop}%
\bibitem [{\citenamefont {Shon}\ and\ \citenamefont {Ando}(1998)}]{q15a}%
  \BibitemOpen
  \bibfield  {author} {\bibinfo {author} {\bibfnamefont {N.}~\bibnamefont
  {Shon}}\ and\ \bibinfo {author} {\bibfnamefont {T.}~\bibnamefont {Ando}},\
  }\href {\doibase 10.1143/JPSJ.67.2421} {\bibfield  {journal} {\bibinfo
  {journal} {J. Phys. Soc. Jpn.}\ }\textbf {\bibinfo {volume} {67}},\ \bibinfo
  {pages} {2421} (\bibinfo {year} {1998})}\BibitemShut {NoStop}%
\bibitem [{\citenamefont {Fukuzawa}\ \emph {et~al.}(2009)\citenamefont
  {Fukuzawa}, \citenamefont {Koshino},\ and\ \citenamefont {Ando}}]{q15b}%
  \BibitemOpen
  \bibfield  {author} {\bibinfo {author} {\bibfnamefont {T.}~\bibnamefont
  {Fukuzawa}}, \bibinfo {author} {\bibfnamefont {M.}~\bibnamefont {Koshino}}, \
  and\ \bibinfo {author} {\bibfnamefont {T.}~\bibnamefont {Ando}},\ }\href
  {\doibase 10.1143/JPSJ.78.094714} {\bibfield  {journal} {\bibinfo  {journal}
  {J. Phys. Soc. Jpn.}\ }\textbf {\bibinfo {volume} {78}},\ \bibinfo {pages}
  {094714} (\bibinfo {year} {2009})}\BibitemShut {NoStop}%
\bibitem [{\citenamefont {Pereira}\ \emph {et~al.}(2006)\citenamefont
  {Pereira}, \citenamefont {Guinea}, \citenamefont {Lopes~dos Santos},
  \citenamefont {Peres},\ and\ \citenamefont {Castro~Neto}}]{q15c}%
  \BibitemOpen
  \bibfield  {author} {\bibinfo {author} {\bibfnamefont {V.~M.}\ \bibnamefont
  {Pereira}}, \bibinfo {author} {\bibfnamefont {F.}~\bibnamefont {Guinea}},
  \bibinfo {author} {\bibfnamefont {J.~M.~B.}\ \bibnamefont {Lopes~dos
  Santos}}, \bibinfo {author} {\bibfnamefont {N.~M.~R.}\ \bibnamefont {Peres}},
  \ and\ \bibinfo {author} {\bibfnamefont {A.~H.}\ \bibnamefont
  {Castro~Neto}},\ }\href {\doibase 10.1103/PhysRevLett.96.036801} {\bibfield
  {journal} {\bibinfo  {journal} {Phys. Rev. Lett.}\ }\textbf {\bibinfo
  {volume} {96}},\ \bibinfo {pages} {036801} (\bibinfo {year}
  {2006})}\BibitemShut {NoStop}%
\bibitem [{\citenamefont {Peres}\ \emph {et~al.}(2006)\citenamefont {Peres},
  \citenamefont {Guinea},\ and\ \citenamefont {Castro~Neto}}]{q16}%
  \BibitemOpen
  \bibfield  {author} {\bibinfo {author} {\bibfnamefont {N.~M.~R.}\
  \bibnamefont {Peres}}, \bibinfo {author} {\bibfnamefont {F.}~\bibnamefont
  {Guinea}}, \ and\ \bibinfo {author} {\bibfnamefont {A.~H.}\ \bibnamefont
  {Castro~Neto}},\ }\href {\doibase 10.1103/PhysRevB.73.125411} {\bibfield
  {journal} {\bibinfo  {journal} {Phys. Rev. B}\ }\textbf {\bibinfo {volume}
  {73}},\ \bibinfo {pages} {125411} (\bibinfo {year} {2006})}\BibitemShut
  {NoStop}%
\bibitem [{\citenamefont {D\'ora}\ \emph {et~al.}(2008)\citenamefont {D\'ora},
  \citenamefont {Ziegler},\ and\ \citenamefont {Thalmeier}}]{q17}%
  \BibitemOpen
  \bibfield  {author} {\bibinfo {author} {\bibfnamefont {B.}~\bibnamefont
  {D\'ora}}, \bibinfo {author} {\bibfnamefont {K.}~\bibnamefont {Ziegler}}, \
  and\ \bibinfo {author} {\bibfnamefont {P.}~\bibnamefont {Thalmeier}},\ }\href
  {\doibase 10.1103/PhysRevB.77.115422} {\bibfield  {journal} {\bibinfo
  {journal} {Phys. Rev. B}\ }\textbf {\bibinfo {volume} {77}},\ \bibinfo
  {pages} {115422} (\bibinfo {year} {2008})}\BibitemShut {NoStop}%
\bibitem [{\citenamefont {Hu}\ \emph {et~al.}(2008)\citenamefont {Hu},
  \citenamefont {Hwang},\ and\ \citenamefont {Das~Sarma}}]{q18}%
  \BibitemOpen
  \bibfield  {author} {\bibinfo {author} {\bibfnamefont {B.~Y.-K.}\
  \bibnamefont {Hu}}, \bibinfo {author} {\bibfnamefont {E.~H.}\ \bibnamefont
  {Hwang}}, \ and\ \bibinfo {author} {\bibfnamefont {S.}~\bibnamefont
  {Das~Sarma}},\ }\href {\doibase 10.1103/PhysRevB.78.165411} {\bibfield
  {journal} {\bibinfo  {journal} {Phys. Rev. B}\ }\textbf {\bibinfo {volume}
  {78}},\ \bibinfo {pages} {165411} (\bibinfo {year} {2008})}\BibitemShut
  {NoStop}%
\bibitem [{\citenamefont {Arimura}\ and\ \citenamefont {Ando}(2012)}]{q25}%
  \BibitemOpen
  \bibfield  {author} {\bibinfo {author} {\bibfnamefont {Y.}~\bibnamefont
  {Arimura}}\ and\ \bibinfo {author} {\bibfnamefont {T.}~\bibnamefont {Ando}},\
  }\href {\doibase 10.1143/JPSJ.81.024702} {\bibfield  {journal} {\bibinfo
  {journal} {J. Phys. Soc. Jpn.}\ }\textbf {\bibinfo {volume} {81}},\ \bibinfo
  {pages} {024702} (\bibinfo {year} {2012})}\BibitemShut {NoStop}%
\bibitem [{\citenamefont {Ando}(2015)}]{q26}%
  \BibitemOpen
  \bibfield  {author} {\bibinfo {author} {\bibfnamefont {T.}~\bibnamefont
  {Ando}},\ }\href {\doibase 10.7566/JPSJ.84.114705} {\bibfield  {journal}
  {\bibinfo  {journal} {J. Phys. Soc. Jpn.}\ }\textbf {\bibinfo {volume}
  {84}},\ \bibinfo {pages} {114705} (\bibinfo {year} {2015})}\BibitemShut
  {NoStop}%
\bibitem [{\citenamefont {Rostami}\ and\ \citenamefont
  {Cappelluti}(2017)}]{q27}%
  \BibitemOpen
  \bibfield  {author} {\bibinfo {author} {\bibfnamefont {H.}~\bibnamefont
  {Rostami}}\ and\ \bibinfo {author} {\bibfnamefont {E.}~\bibnamefont
  {Cappelluti}},\ }\href {\doibase 10.1103/PhysRevB.96.054205} {\bibfield
  {journal} {\bibinfo  {journal} {Phys. Rev. B}\ }\textbf {\bibinfo {volume}
  {96}},\ \bibinfo {pages} {054205} (\bibinfo {year} {2017})}\BibitemShut
  {NoStop}%
\bibitem [{\citenamefont {Balatsky}\ \emph {et~al.}(2006)\citenamefont
  {Balatsky}, \citenamefont {Vekhter},\ and\ \citenamefont {Zhu}}]{q28}%
  \BibitemOpen
  \bibfield  {author} {\bibinfo {author} {\bibfnamefont {A.~V.}\ \bibnamefont
  {Balatsky}}, \bibinfo {author} {\bibfnamefont {I.}~\bibnamefont {Vekhter}}, \
  and\ \bibinfo {author} {\bibfnamefont {J.-X.}\ \bibnamefont {Zhu}},\ }\href
  {\doibase 10.1103/RevModPhys.78.373} {\bibfield  {journal} {\bibinfo
  {journal} {Rev. Mod. Phys.}\ }\textbf {\bibinfo {volume} {78}},\ \bibinfo
  {pages} {373} (\bibinfo {year} {2006})}\BibitemShut {NoStop}%
\bibitem [{\citenamefont {Gonz\'alez}\ and\ \citenamefont
  {Fern\'andez-Rossier}(2012)}]{q29}%
  \BibitemOpen
  \bibfield  {author} {\bibinfo {author} {\bibfnamefont {J.~W.}\ \bibnamefont
  {Gonz\'alez}}\ and\ \bibinfo {author} {\bibfnamefont {J.}~\bibnamefont
  {Fern\'andez-Rossier}},\ }\href {\doibase 10.1103/PhysRevB.86.115327}
  {\bibfield  {journal} {\bibinfo  {journal} {Phys. Rev. B}\ }\textbf {\bibinfo
  {volume} {86}},\ \bibinfo {pages} {115327} (\bibinfo {year}
  {2012})}\BibitemShut {NoStop}%
\bibitem [{\citenamefont {Castro}\ \emph {et~al.}(2015)\citenamefont {Castro},
  \citenamefont {L\'opez-Sancho},\ and\ \citenamefont {Vozmediano}}]{q30}%
  \BibitemOpen
  \bibfield  {author} {\bibinfo {author} {\bibfnamefont {E.~V.}\ \bibnamefont
  {Castro}}, \bibinfo {author} {\bibfnamefont {M.~P.}\ \bibnamefont
  {L\'opez-Sancho}}, \ and\ \bibinfo {author} {\bibfnamefont {M.~A.~H.}\
  \bibnamefont {Vozmediano}},\ }\href {\doibase 10.1103/PhysRevB.92.085410}
  {\bibfield  {journal} {\bibinfo  {journal} {Phys. Rev. B}\ }\textbf {\bibinfo
  {volume} {92}},\ \bibinfo {pages} {085410} (\bibinfo {year}
  {2015})}\BibitemShut {NoStop}%
\bibitem [{\citenamefont {Fu}\ and\ \citenamefont {Kane}(2007)}]{q4}%
  \BibitemOpen
  \bibfield  {author} {\bibinfo {author} {\bibfnamefont {L.}~\bibnamefont
  {Fu}}\ and\ \bibinfo {author} {\bibfnamefont {C.~L.}\ \bibnamefont {Kane}},\
  }\href {\doibase 10.1103/PhysRevB.76.045302} {\bibfield  {journal} {\bibinfo
  {journal} {Phys. Rev. B}\ }\textbf {\bibinfo {volume} {76}},\ \bibinfo
  {pages} {045302} (\bibinfo {year} {2007})}\BibitemShut {NoStop}%
\bibitem [{\citenamefont {Xia}\ \emph {et~al.}(2009)\citenamefont {Xia},
  \citenamefont {Qian}, \citenamefont {Hsieh}, \citenamefont {Wray},
  \citenamefont {Pal}, \citenamefont {Lin}, \citenamefont {Bansil},
  \citenamefont {Grauer}, \citenamefont {Hor}, \citenamefont {Cava},\ and\
  \citenamefont {Hasan}}]{q5}%
  \BibitemOpen
  \bibfield  {author} {\bibinfo {author} {\bibfnamefont {Y.}~\bibnamefont
  {Xia}}, \bibinfo {author} {\bibfnamefont {D.}~\bibnamefont {Qian}}, \bibinfo
  {author} {\bibfnamefont {D.}~\bibnamefont {Hsieh}}, \bibinfo {author}
  {\bibfnamefont {L.}~\bibnamefont {Wray}}, \bibinfo {author} {\bibfnamefont
  {A.}~\bibnamefont {Pal}}, \bibinfo {author} {\bibfnamefont {H.}~\bibnamefont
  {Lin}}, \bibinfo {author} {\bibfnamefont {A.}~\bibnamefont {Bansil}},
  \bibinfo {author} {\bibfnamefont {D.}~\bibnamefont {Grauer}}, \bibinfo
  {author} {\bibfnamefont {Y.~S.}\ \bibnamefont {Hor}}, \bibinfo {author}
  {\bibfnamefont {R.~J.}\ \bibnamefont {Cava}}, \ and\ \bibinfo {author}
  {\bibfnamefont {M.~Z.}\ \bibnamefont {Hasan}},\ }\href {\doibase
  10.1038/nphys1274} {\bibfield  {journal} {\bibinfo  {journal} {Nature Phys.}\
  }\textbf {\bibinfo {volume} {5}},\ \bibinfo {pages} {398} (\bibinfo {year}
  {2009})}\BibitemShut {NoStop}%
\bibitem [{\citenamefont {Zhang}\ \emph {et~al.}(2009)\citenamefont {Zhang},
  \citenamefont {Liu}, \citenamefont {Qi}, \citenamefont {Dai}, \citenamefont
  {Fang},\ and\ \citenamefont {Zhang}}]{q6}%
  \BibitemOpen
  \bibfield  {author} {\bibinfo {author} {\bibfnamefont {H.}~\bibnamefont
  {Zhang}}, \bibinfo {author} {\bibfnamefont {C.-X.}\ \bibnamefont {Liu}},
  \bibinfo {author} {\bibfnamefont {X.-L.}\ \bibnamefont {Qi}}, \bibinfo
  {author} {\bibfnamefont {X.}~\bibnamefont {Dai}}, \bibinfo {author}
  {\bibfnamefont {Z.}~\bibnamefont {Fang}}, \ and\ \bibinfo {author}
  {\bibfnamefont {S.-C.}\ \bibnamefont {Zhang}},\ }\href {\doibase
  10.1038/nphys1270} {\bibfield  {journal} {\bibinfo  {journal} {Nature Phys.}\
  }\textbf {\bibinfo {volume} {5}},\ \bibinfo {pages} {438} (\bibinfo {year}
  {2009})}\BibitemShut {NoStop}%
\bibitem [{\citenamefont {Shan}\ \emph {et~al.}(2010)\citenamefont {Shan},
  \citenamefont {Lu},\ and\ \citenamefont {Shen}}]{q31}%
  \BibitemOpen
  \bibfield  {author} {\bibinfo {author} {\bibfnamefont {W.-Y.}\ \bibnamefont
  {Shan}}, \bibinfo {author} {\bibfnamefont {H.-Z.}\ \bibnamefont {Lu}}, \ and\
  \bibinfo {author} {\bibfnamefont {S.-Q.}\ \bibnamefont {Shen}},\ }\href
  {\doibase 10.1088/1367-2630/12/4/043048} {\bibfield  {journal} {\bibinfo
  {journal} {New J. Phys.}\ }\textbf {\bibinfo {volume} {12}},\ \bibinfo
  {pages} {043048} (\bibinfo {year} {2010})}\BibitemShut {NoStop}%
\bibitem [{\citenamefont {Lu}\ \emph {et~al.}(2010)\citenamefont {Lu},
  \citenamefont {Shan}, \citenamefont {Yao}, \citenamefont {Niu},\ and\
  \citenamefont {Shen}}]{q32}%
  \BibitemOpen
  \bibfield  {author} {\bibinfo {author} {\bibfnamefont {H.-Z.}\ \bibnamefont
  {Lu}}, \bibinfo {author} {\bibfnamefont {W.-Y.}\ \bibnamefont {Shan}},
  \bibinfo {author} {\bibfnamefont {W.}~\bibnamefont {Yao}}, \bibinfo {author}
  {\bibfnamefont {Q.}~\bibnamefont {Niu}}, \ and\ \bibinfo {author}
  {\bibfnamefont {S.-Q.}\ \bibnamefont {Shen}},\ }\href {\doibase
  10.1103/PhysRevB.81.115407} {\bibfield  {journal} {\bibinfo  {journal} {Phys.
  Rev. B}\ }\textbf {\bibinfo {volume} {81}},\ \bibinfo {pages} {115407}
  (\bibinfo {year} {2010})}\BibitemShut {NoStop}%
\bibitem [{\citenamefont {Liu}\ \emph {et~al.}(2010)\citenamefont {Liu},
  \citenamefont {Zhang}, \citenamefont {Yan}, \citenamefont {Qi}, \citenamefont
  {Frauenheim}, \citenamefont {Dai}, \citenamefont {Fang},\ and\ \citenamefont
  {Zhang}}]{q33}%
  \BibitemOpen
  \bibfield  {author} {\bibinfo {author} {\bibfnamefont {C.-X.}\ \bibnamefont
  {Liu}}, \bibinfo {author} {\bibfnamefont {H.}~\bibnamefont {Zhang}}, \bibinfo
  {author} {\bibfnamefont {B.}~\bibnamefont {Yan}}, \bibinfo {author}
  {\bibfnamefont {X.-L.}\ \bibnamefont {Qi}}, \bibinfo {author} {\bibfnamefont
  {T.}~\bibnamefont {Frauenheim}}, \bibinfo {author} {\bibfnamefont
  {X.}~\bibnamefont {Dai}}, \bibinfo {author} {\bibfnamefont {Z.}~\bibnamefont
  {Fang}}, \ and\ \bibinfo {author} {\bibfnamefont {S.-C.}\ \bibnamefont
  {Zhang}},\ }\href {\doibase 10.1103/PhysRevB.81.041307} {\bibfield  {journal}
  {\bibinfo  {journal} {Phys. Rev. B}\ }\textbf {\bibinfo {volume} {81}},\
  \bibinfo {pages} {041307} (\bibinfo {year} {2010})}\BibitemShut {NoStop}%
\bibitem [{\citenamefont {Bernevig}\ \emph {et~al.}(2006)\citenamefont
  {Bernevig}, \citenamefont {Hughes},\ and\ \citenamefont {Zhang}}]{q9}%
  \BibitemOpen
  \bibfield  {author} {\bibinfo {author} {\bibfnamefont {B.~A.}\ \bibnamefont
  {Bernevig}}, \bibinfo {author} {\bibfnamefont {T.~L.}\ \bibnamefont
  {Hughes}}, \ and\ \bibinfo {author} {\bibfnamefont {S.-C.}\ \bibnamefont
  {Zhang}},\ }\href {\doibase 10.1126/science.1133734} {\bibfield  {journal}
  {\bibinfo  {journal} {Science}\ }\textbf {\bibinfo {volume} {314}},\ \bibinfo
  {pages} {1757} (\bibinfo {year} {2006})}\BibitemShut {NoStop}%
\bibitem [{\citenamefont {K\"{o}nig}\ \emph {et~al.}(2007)\citenamefont
  {K\"{o}nig}, \citenamefont {Wiedmann}, \citenamefont {Br\"{u}ne},
  \citenamefont {Roth}, \citenamefont {Buhmann}, \citenamefont {Molenkamp},
  \citenamefont {Qi},\ and\ \citenamefont {Zhang}}]{q10}%
  \BibitemOpen
  \bibfield  {author} {\bibinfo {author} {\bibfnamefont {M.}~\bibnamefont
  {K\"{o}nig}}, \bibinfo {author} {\bibfnamefont {S.}~\bibnamefont {Wiedmann}},
  \bibinfo {author} {\bibfnamefont {C.}~\bibnamefont {Br\"{u}ne}}, \bibinfo
  {author} {\bibfnamefont {A.}~\bibnamefont {Roth}}, \bibinfo {author}
  {\bibfnamefont {H.}~\bibnamefont {Buhmann}}, \bibinfo {author} {\bibfnamefont
  {L.~W.}\ \bibnamefont {Molenkamp}}, \bibinfo {author} {\bibfnamefont {X.-L.}\
  \bibnamefont {Qi}}, \ and\ \bibinfo {author} {\bibfnamefont {S.-C.}\
  \bibnamefont {Zhang}},\ }\href {\doibase 10.1126/science.1148047} {\bibfield
  {journal} {\bibinfo  {journal} {Science}\ }\textbf {\bibinfo {volume}
  {318}},\ \bibinfo {pages} {766} (\bibinfo {year} {2007})}\BibitemShut
  {NoStop}%
\bibitem [{\citenamefont {B\"{u}ttner}\ \emph {et~al.}(2011)\citenamefont
  {B\"{u}ttner}, \citenamefont {Liu}, \citenamefont {Tkachov}, \citenamefont
  {Novik}, \citenamefont {Br\"{u}ne}, \citenamefont {Buhmann}, \citenamefont
  {Hankiewicz}, \citenamefont {Recher}, \citenamefont {Trauzettel},
  \citenamefont {Zhang},\ and\ \citenamefont {Molenkamp}}]{q11}%
  \BibitemOpen
  \bibfield  {author} {\bibinfo {author} {\bibfnamefont {B.}~\bibnamefont
  {B\"{u}ttner}}, \bibinfo {author} {\bibfnamefont {C.}~\bibnamefont {Liu}},
  \bibinfo {author} {\bibfnamefont {G.}~\bibnamefont {Tkachov}}, \bibinfo
  {author} {\bibfnamefont {E.}~\bibnamefont {Novik}}, \bibinfo {author}
  {\bibfnamefont {C.}~\bibnamefont {Br\"{u}ne}}, \bibinfo {author}
  {\bibfnamefont {H.}~\bibnamefont {Buhmann}}, \bibinfo {author} {\bibfnamefont
  {E.}~\bibnamefont {Hankiewicz}}, \bibinfo {author} {\bibfnamefont
  {P.}~\bibnamefont {Recher}}, \bibinfo {author} {\bibfnamefont
  {B.}~\bibnamefont {Trauzettel}}, \bibinfo {author} {\bibfnamefont
  {S.}~\bibnamefont {Zhang}}, \ and\ \bibinfo {author} {\bibfnamefont
  {L.}~\bibnamefont {Molenkamp}},\ }\href {\doibase 10.1038/nphys1914}
  {\bibfield  {journal} {\bibinfo  {journal} {Nat. Phys.}\ }\textbf {\bibinfo
  {volume} {7}},\ \bibinfo {pages} {418} (\bibinfo {year} {2011})}\BibitemShut
  {NoStop}%
\bibitem [{\citenamefont {Krishtopenko}\ and\ \citenamefont
  {Teppe}(2018{\natexlab{a}})}]{q12}%
  \BibitemOpen
  \bibfield  {author} {\bibinfo {author} {\bibfnamefont {S.~S.}\ \bibnamefont
  {Krishtopenko}}\ and\ \bibinfo {author} {\bibfnamefont {F.}~\bibnamefont
  {Teppe}},\ }\href {\doibase 10.1126/sciadv.aap7529} {\bibfield  {journal}
  {\bibinfo  {journal} {Sci. Adv.}\ }\textbf {\bibinfo {volume} {4}},\ \bibinfo
  {pages} {eaap7529} (\bibinfo {year} {2018}{\natexlab{a}})}\BibitemShut
  {NoStop}%
\bibitem [{\citenamefont {Krishtopenko}\ \emph {et~al.}(2018)\citenamefont
  {Krishtopenko}, \citenamefont {Ruffenach}, \citenamefont {Gonzalez-Posada},
  \citenamefont {Boissier}, \citenamefont {Marcinkiewicz}, \citenamefont
  {Fadeev}, \citenamefont {Kadykov}, \citenamefont {Rumyantsev}, \citenamefont
  {Morozov}, \citenamefont {Gavrilenko}, \citenamefont {Consejo}, \citenamefont
  {Desrat}, \citenamefont {Jouault}, \citenamefont {Knap}, \citenamefont
  {Tourni\'e},\ and\ \citenamefont {Teppe}}]{q13}%
  \BibitemOpen
  \bibfield  {author} {\bibinfo {author} {\bibfnamefont {S.~S.}\ \bibnamefont
  {Krishtopenko}}, \bibinfo {author} {\bibfnamefont {S.}~\bibnamefont
  {Ruffenach}}, \bibinfo {author} {\bibfnamefont {F.}~\bibnamefont
  {Gonzalez-Posada}}, \bibinfo {author} {\bibfnamefont {G.}~\bibnamefont
  {Boissier}}, \bibinfo {author} {\bibfnamefont {M.}~\bibnamefont
  {Marcinkiewicz}}, \bibinfo {author} {\bibfnamefont {M.~A.}\ \bibnamefont
  {Fadeev}}, \bibinfo {author} {\bibfnamefont {A.~M.}\ \bibnamefont {Kadykov}},
  \bibinfo {author} {\bibfnamefont {V.~V.}\ \bibnamefont {Rumyantsev}},
  \bibinfo {author} {\bibfnamefont {S.~V.}\ \bibnamefont {Morozov}}, \bibinfo
  {author} {\bibfnamefont {V.~I.}\ \bibnamefont {Gavrilenko}}, \bibinfo
  {author} {\bibfnamefont {C.}~\bibnamefont {Consejo}}, \bibinfo {author}
  {\bibfnamefont {W.}~\bibnamefont {Desrat}}, \bibinfo {author} {\bibfnamefont
  {B.}~\bibnamefont {Jouault}}, \bibinfo {author} {\bibfnamefont
  {W.}~\bibnamefont {Knap}}, \bibinfo {author} {\bibfnamefont {E.}~\bibnamefont
  {Tourni\'e}}, \ and\ \bibinfo {author} {\bibfnamefont {F.}~\bibnamefont
  {Teppe}},\ }\href {\doibase 10.1103/PhysRevB.97.245419} {\bibfield  {journal}
  {\bibinfo  {journal} {Phys. Rev. B}\ }\textbf {\bibinfo {volume} {97}},\
  \bibinfo {pages} {245419} (\bibinfo {year} {2018})}\BibitemShut {NoStop}%
\bibitem [{\citenamefont {Krishtopenko}\ \emph {et~al.}(2019)\citenamefont
  {Krishtopenko}, \citenamefont {Desrat}, \citenamefont {Spirin}, \citenamefont
  {Consejo}, \citenamefont {Ruffenach}, \citenamefont {Gonzalez-Posada},
  \citenamefont {Jouault}, \citenamefont {Knap}, \citenamefont {Maremyanin},
  \citenamefont {Gavrilenko}, \citenamefont {Boissier}, \citenamefont {Torres},
  \citenamefont {Zaknoune}, \citenamefont {Tourni\'e},\ and\ \citenamefont
  {Teppe}}]{q14}%
  \BibitemOpen
  \bibfield  {author} {\bibinfo {author} {\bibfnamefont {S.~S.}\ \bibnamefont
  {Krishtopenko}}, \bibinfo {author} {\bibfnamefont {W.}~\bibnamefont
  {Desrat}}, \bibinfo {author} {\bibfnamefont {K.~E.}\ \bibnamefont {Spirin}},
  \bibinfo {author} {\bibfnamefont {C.}~\bibnamefont {Consejo}}, \bibinfo
  {author} {\bibfnamefont {S.}~\bibnamefont {Ruffenach}}, \bibinfo {author}
  {\bibfnamefont {F.}~\bibnamefont {Gonzalez-Posada}}, \bibinfo {author}
  {\bibfnamefont {B.}~\bibnamefont {Jouault}}, \bibinfo {author} {\bibfnamefont
  {W.}~\bibnamefont {Knap}}, \bibinfo {author} {\bibfnamefont {K.~V.}\
  \bibnamefont {Maremyanin}}, \bibinfo {author} {\bibfnamefont {V.~I.}\
  \bibnamefont {Gavrilenko}}, \bibinfo {author} {\bibfnamefont
  {G.}~\bibnamefont {Boissier}}, \bibinfo {author} {\bibfnamefont
  {J.}~\bibnamefont {Torres}}, \bibinfo {author} {\bibfnamefont
  {M.}~\bibnamefont {Zaknoune}}, \bibinfo {author} {\bibfnamefont
  {E.}~\bibnamefont {Tourni\'e}}, \ and\ \bibinfo {author} {\bibfnamefont
  {F.}~\bibnamefont {Teppe}},\ }\href {\doibase 10.1103/PhysRevB.99.121405}
  {\bibfield  {journal} {\bibinfo  {journal} {Phys. Rev. B}\ }\textbf {\bibinfo
  {volume} {99}},\ \bibinfo {pages} {121405} (\bibinfo {year}
  {2019})}\BibitemShut {NoStop}%
\bibitem [{\citenamefont {Shen}\ \emph {et~al.}(2011)\citenamefont {Shen},
  \citenamefont {Shan},\ and\ \citenamefont {Lu}}]{q47}%
  \BibitemOpen
  \bibfield  {author} {\bibinfo {author} {\bibfnamefont {S.-Q.}\ \bibnamefont
  {Shen}}, \bibinfo {author} {\bibfnamefont {W.-Y.}\ \bibnamefont {Shan}}, \
  and\ \bibinfo {author} {\bibfnamefont {H.-Z.}\ \bibnamefont {Lu}},\ }\href
  {\doibase 10.1142/S2010324711000057} {\bibfield  {journal} {\bibinfo
  {journal} {SPIN}\ }\textbf {\bibinfo {volume} {01}},\ \bibinfo {pages} {33}
  (\bibinfo {year} {2011})}\BibitemShut {NoStop}%
\bibitem [{\citenamefont {Li}\ \emph {et~al.}(2009)\citenamefont {Li},
  \citenamefont {Chu}, \citenamefont {Jain},\ and\ \citenamefont {Shen}}]{q34}%
  \BibitemOpen
  \bibfield  {author} {\bibinfo {author} {\bibfnamefont {J.}~\bibnamefont
  {Li}}, \bibinfo {author} {\bibfnamefont {R.-L.}\ \bibnamefont {Chu}},
  \bibinfo {author} {\bibfnamefont {J.~K.}\ \bibnamefont {Jain}}, \ and\
  \bibinfo {author} {\bibfnamefont {S.-Q.}\ \bibnamefont {Shen}},\ }\href
  {\doibase 10.1103/PhysRevLett.102.136806} {\bibfield  {journal} {\bibinfo
  {journal} {Phys. Rev. Lett.}\ }\textbf {\bibinfo {volume} {102}},\ \bibinfo
  {pages} {136806} (\bibinfo {year} {2009})}\BibitemShut {NoStop}%
\bibitem [{\citenamefont {Groth}\ \emph {et~al.}(2009)\citenamefont {Groth},
  \citenamefont {Wimmer}, \citenamefont {Akhmerov}, \citenamefont
  {Tworzyd\l{}o},\ and\ \citenamefont {Beenakker}}]{q35}%
  \BibitemOpen
  \bibfield  {author} {\bibinfo {author} {\bibfnamefont {C.~W.}\ \bibnamefont
  {Groth}}, \bibinfo {author} {\bibfnamefont {M.}~\bibnamefont {Wimmer}},
  \bibinfo {author} {\bibfnamefont {A.~R.}\ \bibnamefont {Akhmerov}}, \bibinfo
  {author} {\bibfnamefont {J.}~\bibnamefont {Tworzyd\l{}o}}, \ and\ \bibinfo
  {author} {\bibfnamefont {C.~W.~J.}\ \bibnamefont {Beenakker}},\ }\href
  {\doibase 10.1103/PhysRevLett.103.196805} {\bibfield  {journal} {\bibinfo
  {journal} {Phys. Rev. Lett.}\ }\textbf {\bibinfo {volume} {103}},\ \bibinfo
  {pages} {196805} (\bibinfo {year} {2009})}\BibitemShut {NoStop}%
\bibitem [{\citenamefont {Chen}\ \emph {et~al.}(2012)\citenamefont {Chen},
  \citenamefont {Liu}, \citenamefont {Lin}, \citenamefont {Zhang},\ and\
  \citenamefont {Jiang}}]{q36}%
  \BibitemOpen
  \bibfield  {author} {\bibinfo {author} {\bibfnamefont {L.}~\bibnamefont
  {Chen}}, \bibinfo {author} {\bibfnamefont {Q.}~\bibnamefont {Liu}}, \bibinfo
  {author} {\bibfnamefont {X.}~\bibnamefont {Lin}}, \bibinfo {author}
  {\bibfnamefont {X.}~\bibnamefont {Zhang}}, \ and\ \bibinfo {author}
  {\bibfnamefont {X.}~\bibnamefont {Jiang}},\ }\href {\doibase
  10.1088/1367-2630/14/4/043028} {\bibfield  {journal} {\bibinfo  {journal}
  {New J. of Phys.}\ }\textbf {\bibinfo {volume} {14}},\ \bibinfo {pages}
  {043028} (\bibinfo {year} {2012})}\BibitemShut {NoStop}%
\bibitem [{\citenamefont {Girschik}\ \emph {et~al.}(2013)\citenamefont
  {Girschik}, \citenamefont {Libisch},\ and\ \citenamefont {Rotter}}]{q37}%
  \BibitemOpen
  \bibfield  {author} {\bibinfo {author} {\bibfnamefont {A.}~\bibnamefont
  {Girschik}}, \bibinfo {author} {\bibfnamefont {F.}~\bibnamefont {Libisch}}, \
  and\ \bibinfo {author} {\bibfnamefont {S.}~\bibnamefont {Rotter}},\ }\href
  {\doibase 10.1103/PhysRevB.88.014201} {\bibfield  {journal} {\bibinfo
  {journal} {Phys. Rev. B}\ }\textbf {\bibinfo {volume} {88}},\ \bibinfo
  {pages} {014201} (\bibinfo {year} {2013})}\BibitemShut {NoStop}%
\bibitem [{\citenamefont {Krishtopenko}\ and\ \citenamefont
  {Teppe}(2018{\natexlab{b}})}]{q38}%
  \BibitemOpen
  \bibfield  {author} {\bibinfo {author} {\bibfnamefont {S.~S.}\ \bibnamefont
  {Krishtopenko}}\ and\ \bibinfo {author} {\bibfnamefont {F.}~\bibnamefont
  {Teppe}},\ }\href {\doibase 10.1103/PhysRevB.97.165408} {\bibfield  {journal}
  {\bibinfo  {journal} {Phys. Rev. B}\ }\textbf {\bibinfo {volume} {97}},\
  \bibinfo {pages} {165408} (\bibinfo {year} {2018}{\natexlab{b}})}\BibitemShut
  {NoStop}%
\bibitem [{\citenamefont {K\"{o}nig}\ \emph {et~al.}(2008)\citenamefont
  {K\"{o}nig}, \citenamefont {Buhmann}, \citenamefont {W.~Molenkamp},
  \citenamefont {Hughes}, \citenamefont {Liu}, \citenamefont {Qi},\ and\
  \citenamefont {Zhang}}]{q39}%
  \BibitemOpen
  \bibfield  {author} {\bibinfo {author} {\bibfnamefont {M.}~\bibnamefont
  {K\"{o}nig}}, \bibinfo {author} {\bibfnamefont {H.}~\bibnamefont {Buhmann}},
  \bibinfo {author} {\bibfnamefont {L.}~\bibnamefont {W.~Molenkamp}}, \bibinfo
  {author} {\bibfnamefont {T.}~\bibnamefont {Hughes}}, \bibinfo {author}
  {\bibfnamefont {C.-X.}\ \bibnamefont {Liu}}, \bibinfo {author} {\bibfnamefont
  {X.-L.}\ \bibnamefont {Qi}}, \ and\ \bibinfo {author} {\bibfnamefont {S.-C.}\
  \bibnamefont {Zhang}},\ }\href {\doibase 10.1143/JPSJ.77.031007} {\bibfield
  {journal} {\bibinfo  {journal} {J. Phys. Soc. Jpn.}\ }\textbf {\bibinfo
  {volume} {77}},\ \bibinfo {pages} {031007} (\bibinfo {year}
  {2008})}\BibitemShut {NoStop}%
\bibitem [{\citenamefont {Rothe}\ \emph {et~al.}(2010)\citenamefont {Rothe},
  \citenamefont {Reinthaler}, \citenamefont {Liu}, \citenamefont {Molenkamp},
  \citenamefont {Zhang},\ and\ \citenamefont {Hankiewicz}}]{q40}%
  \BibitemOpen
  \bibfield  {author} {\bibinfo {author} {\bibfnamefont {D.~G.}\ \bibnamefont
  {Rothe}}, \bibinfo {author} {\bibfnamefont {R.~W.}\ \bibnamefont
  {Reinthaler}}, \bibinfo {author} {\bibfnamefont {C.-X.}\ \bibnamefont {Liu}},
  \bibinfo {author} {\bibfnamefont {L.~W.}\ \bibnamefont {Molenkamp}}, \bibinfo
  {author} {\bibfnamefont {S.-C.}\ \bibnamefont {Zhang}}, \ and\ \bibinfo
  {author} {\bibfnamefont {E.~M.}\ \bibnamefont {Hankiewicz}},\ }\href
  {\doibase 10.1088/1367-2630/12/6/065012} {\bibfield  {journal} {\bibinfo
  {journal} {New J. Phys.}\ }\textbf {\bibinfo {volume} {12}},\ \bibinfo
  {pages} {065012} (\bibinfo {year} {2010})}\BibitemShut {NoStop}%
\bibitem [{SM()}]{SM}%
  \BibitemOpen
  \href@noop {} {\bibinfo  {journal} {See Supplemental Materials, which also
  contains Ref.~[53], for any details of the SCBA within the two-band BHZ model
  [28] and the four-band advanced 2D Hamiltonian [39]. The mobility calculation
  for the short-range impurities based on the BHZ Hamiltonian, as well as the
  parameters for both models are also provided therein}\ }\BibitemShut
  {NoStop}%
\bibitem [{\citenamefont {Kadykov}\ \emph {et~al.}(2018)\citenamefont
  {Kadykov}, \citenamefont {Krishtopenko}, \citenamefont {Jouault},
  \citenamefont {Desrat}, \citenamefont {Knap}, \citenamefont {Ruffenach},
  \citenamefont {Consejo}, \citenamefont {Torres}, \citenamefont {Morozov},
  \citenamefont {Mikhailov}, \citenamefont {Dvoretskii},\ and\ \citenamefont
  {Teppe}}]{q41}%
  \BibitemOpen
\bibfield  {journal} {  }\bibfield  {author} {\bibinfo {author} {\bibfnamefont
  {A.~M.}\ \bibnamefont {Kadykov}}, \bibinfo {author} {\bibfnamefont {S.~S.}\
  \bibnamefont {Krishtopenko}}, \bibinfo {author} {\bibfnamefont
  {B.}~\bibnamefont {Jouault}}, \bibinfo {author} {\bibfnamefont
  {W.}~\bibnamefont {Desrat}}, \bibinfo {author} {\bibfnamefont
  {W.}~\bibnamefont {Knap}}, \bibinfo {author} {\bibfnamefont {S.}~\bibnamefont
  {Ruffenach}}, \bibinfo {author} {\bibfnamefont {C.}~\bibnamefont {Consejo}},
  \bibinfo {author} {\bibfnamefont {J.}~\bibnamefont {Torres}}, \bibinfo
  {author} {\bibfnamefont {S.~V.}\ \bibnamefont {Morozov}}, \bibinfo {author}
  {\bibfnamefont {N.~N.}\ \bibnamefont {Mikhailov}}, \bibinfo {author}
  {\bibfnamefont {S.~A.}\ \bibnamefont {Dvoretskii}}, \ and\ \bibinfo {author}
  {\bibfnamefont {F.}~\bibnamefont {Teppe}},\ }\href {\doibase
  10.1103/PhysRevLett.120.086401} {\bibfield  {journal} {\bibinfo  {journal}
  {Phys. Rev. Lett.}\ }\textbf {\bibinfo {volume} {120}},\ \bibinfo {pages}
  {086401} (\bibinfo {year} {2018})}\BibitemShut {NoStop}%
\bibitem [{\citenamefont {Krishtopenko}\ \emph
  {et~al.}(2016{\natexlab{a}})\citenamefont {Krishtopenko}, \citenamefont
  {Yahniuk}, \citenamefont {But}, \citenamefont {Gavrilenko}, \citenamefont
  {Knap},\ and\ \citenamefont {Teppe}}]{q42}%
  \BibitemOpen
  \bibfield  {author} {\bibinfo {author} {\bibfnamefont {S.~S.}\ \bibnamefont
  {Krishtopenko}}, \bibinfo {author} {\bibfnamefont {I.}~\bibnamefont
  {Yahniuk}}, \bibinfo {author} {\bibfnamefont {D.~B.}\ \bibnamefont {But}},
  \bibinfo {author} {\bibfnamefont {V.~I.}\ \bibnamefont {Gavrilenko}},
  \bibinfo {author} {\bibfnamefont {W.}~\bibnamefont {Knap}}, \ and\ \bibinfo
  {author} {\bibfnamefont {F.}~\bibnamefont {Teppe}},\ }\href {\doibase
  10.1103/PhysRevB.94.245402} {\bibfield  {journal} {\bibinfo  {journal} {Phys.
  Rev. B}\ }\textbf {\bibinfo {volume} {94}},\ \bibinfo {pages} {245402}
  (\bibinfo {year} {2016}{\natexlab{a}})}\BibitemShut {NoStop}%
\bibitem [{\citenamefont {Wiedmann}\ \emph {et~al.}(2015)\citenamefont
  {Wiedmann}, \citenamefont {Jost}, \citenamefont {Thienel}, \citenamefont
  {Br\"une}, \citenamefont {Leubner}, \citenamefont {Buhmann}, \citenamefont
  {Molenkamp}, \citenamefont {Maan},\ and\ \citenamefont {Zeitler}}]{q43}%
  \BibitemOpen
  \bibfield  {author} {\bibinfo {author} {\bibfnamefont {S.}~\bibnamefont
  {Wiedmann}}, \bibinfo {author} {\bibfnamefont {A.}~\bibnamefont {Jost}},
  \bibinfo {author} {\bibfnamefont {C.}~\bibnamefont {Thienel}}, \bibinfo
  {author} {\bibfnamefont {C.}~\bibnamefont {Br\"une}}, \bibinfo {author}
  {\bibfnamefont {P.}~\bibnamefont {Leubner}}, \bibinfo {author} {\bibfnamefont
  {H.}~\bibnamefont {Buhmann}}, \bibinfo {author} {\bibfnamefont {L.~W.}\
  \bibnamefont {Molenkamp}}, \bibinfo {author} {\bibfnamefont {J.~C.}\
  \bibnamefont {Maan}}, \ and\ \bibinfo {author} {\bibfnamefont
  {U.}~\bibnamefont {Zeitler}},\ }\href {\doibase 10.1103/PhysRevB.91.205311}
  {\bibfield  {journal} {\bibinfo  {journal} {Phys. Rev. B}\ }\textbf {\bibinfo
  {volume} {91}},\ \bibinfo {pages} {205311} (\bibinfo {year}
  {2015})}\BibitemShut {NoStop}%
\bibitem [{\citenamefont {Marcinkiewicz}\ \emph {et~al.}(2017)\citenamefont
  {Marcinkiewicz}, \citenamefont {Ruffenach}, \citenamefont {Krishtopenko},
  \citenamefont {Kadykov}, \citenamefont {Consejo}, \citenamefont {But},
  \citenamefont {Desrat}, \citenamefont {Knap}, \citenamefont {Torres},
  \citenamefont {Ikonnikov}, \citenamefont {Spirin}, \citenamefont {Morozov},
  \citenamefont {Gavrilenko}, \citenamefont {Mikhailov}, \citenamefont
  {Dvoretskii},\ and\ \citenamefont {Teppe}}]{q44}%
  \BibitemOpen
  \bibfield  {author} {\bibinfo {author} {\bibfnamefont {M.}~\bibnamefont
  {Marcinkiewicz}}, \bibinfo {author} {\bibfnamefont {S.}~\bibnamefont
  {Ruffenach}}, \bibinfo {author} {\bibfnamefont {S.~S.}\ \bibnamefont
  {Krishtopenko}}, \bibinfo {author} {\bibfnamefont {A.~M.}\ \bibnamefont
  {Kadykov}}, \bibinfo {author} {\bibfnamefont {C.}~\bibnamefont {Consejo}},
  \bibinfo {author} {\bibfnamefont {D.~B.}\ \bibnamefont {But}}, \bibinfo
  {author} {\bibfnamefont {W.}~\bibnamefont {Desrat}}, \bibinfo {author}
  {\bibfnamefont {W.}~\bibnamefont {Knap}}, \bibinfo {author} {\bibfnamefont
  {J.}~\bibnamefont {Torres}}, \bibinfo {author} {\bibfnamefont {A.~V.}\
  \bibnamefont {Ikonnikov}}, \bibinfo {author} {\bibfnamefont {K.~E.}\
  \bibnamefont {Spirin}}, \bibinfo {author} {\bibfnamefont {S.~V.}\
  \bibnamefont {Morozov}}, \bibinfo {author} {\bibfnamefont {V.~I.}\
  \bibnamefont {Gavrilenko}}, \bibinfo {author} {\bibfnamefont {N.~N.}\
  \bibnamefont {Mikhailov}}, \bibinfo {author} {\bibfnamefont {S.~A.}\
  \bibnamefont {Dvoretskii}}, \ and\ \bibinfo {author} {\bibfnamefont
  {F.}~\bibnamefont {Teppe}},\ }\href {\doibase 10.1103/PhysRevB.96.035405}
  {\bibfield  {journal} {\bibinfo  {journal} {Phys. Rev. B}\ }\textbf {\bibinfo
  {volume} {96}},\ \bibinfo {pages} {035405} (\bibinfo {year}
  {2017})}\BibitemShut {NoStop}%
\bibitem [{\citenamefont {Leubner}\ \emph {et~al.}(2016)\citenamefont
  {Leubner}, \citenamefont {Lunczer}, \citenamefont {Br\"une}, \citenamefont
  {Buhmann},\ and\ \citenamefont {Molenkamp}}]{q45}%
  \BibitemOpen
  \bibfield  {author} {\bibinfo {author} {\bibfnamefont {P.}~\bibnamefont
  {Leubner}}, \bibinfo {author} {\bibfnamefont {L.}~\bibnamefont {Lunczer}},
  \bibinfo {author} {\bibfnamefont {C.}~\bibnamefont {Br\"une}}, \bibinfo
  {author} {\bibfnamefont {H.}~\bibnamefont {Buhmann}}, \ and\ \bibinfo
  {author} {\bibfnamefont {L.~W.}\ \bibnamefont {Molenkamp}},\ }\href {\doibase
  10.1103/PhysRevLett.117.086403} {\bibfield  {journal} {\bibinfo  {journal}
  {Phys. Rev. Lett.}\ }\textbf {\bibinfo {volume} {117}},\ \bibinfo {pages}
  {086403} (\bibinfo {year} {2016})}\BibitemShut {NoStop}%
\bibitem [{\citenamefont {Yahniuk}\ \emph {et~al.}(2019)\citenamefont
  {Yahniuk}, \citenamefont {Krishtopenko}, \citenamefont {Grabecki},
  \citenamefont {Jouault}, \citenamefont {Consejo}, \citenamefont {Desrat},
  \citenamefont {Majewicz}, \citenamefont {Kadykov}, \citenamefont {Spirin},
  \citenamefont {Gavrilenko}, \citenamefont {Mikhailov}, \citenamefont
  {Dvoretsky}, \citenamefont {But}, \citenamefont {Teppe}, \citenamefont
  {Wrobel}, \citenamefont {Cywinski}, \citenamefont {Kret}, \citenamefont
  {Dietl},\ and\ \citenamefont {Knap}}]{q46}%
  \BibitemOpen
  \bibfield  {author} {\bibinfo {author} {\bibfnamefont {I.}~\bibnamefont
  {Yahniuk}}, \bibinfo {author} {\bibfnamefont {S.~S.}\ \bibnamefont
  {Krishtopenko}}, \bibinfo {author} {\bibfnamefont {G.}~\bibnamefont
  {Grabecki}}, \bibinfo {author} {\bibfnamefont {B.}~\bibnamefont {Jouault}},
  \bibinfo {author} {\bibfnamefont {C.}~\bibnamefont {Consejo}}, \bibinfo
  {author} {\bibfnamefont {W.}~\bibnamefont {Desrat}}, \bibinfo {author}
  {\bibfnamefont {M.}~\bibnamefont {Majewicz}}, \bibinfo {author}
  {\bibfnamefont {A.~M.}\ \bibnamefont {Kadykov}}, \bibinfo {author}
  {\bibfnamefont {K.~E.}\ \bibnamefont {Spirin}}, \bibinfo {author}
  {\bibfnamefont {V.~I.}\ \bibnamefont {Gavrilenko}}, \bibinfo {author}
  {\bibfnamefont {N.~N.}\ \bibnamefont {Mikhailov}}, \bibinfo {author}
  {\bibfnamefont {S.~A.}\ \bibnamefont {Dvoretsky}}, \bibinfo {author}
  {\bibfnamefont {D.~B.}\ \bibnamefont {But}}, \bibinfo {author} {\bibfnamefont
  {F.}~\bibnamefont {Teppe}}, \bibinfo {author} {\bibfnamefont
  {J.}~\bibnamefont {Wrobel}}, \bibinfo {author} {\bibfnamefont
  {G.}~\bibnamefont {Cywinski}}, \bibinfo {author} {\bibfnamefont
  {S.}~\bibnamefont {Kret}}, \bibinfo {author} {\bibfnamefont {T.}~\bibnamefont
  {Dietl}}, \ and\ \bibinfo {author} {\bibfnamefont {W.}~\bibnamefont {Knap}},\
  }\href {\doibase 10.1038/s41535-019-0154-3} {\bibfield  {journal} {\bibinfo
  {journal} {npj Quantum Mater.}\ }\textbf {\bibinfo {volume} {4}},\ \bibinfo
  {pages} {13} (\bibinfo {year} {2019})}\BibitemShut {NoStop}%
\bibitem [{\citenamefont {Kvon}\ \emph {et~al.}(2008)\citenamefont {Kvon},
  \citenamefont {Olshanetsky}, \citenamefont {Kozlov}, \citenamefont
  {Mikhailov},\ and\ \citenamefont {Dvoretskii}}]{q48}%
  \BibitemOpen
  \bibfield  {author} {\bibinfo {author} {\bibfnamefont {Z.~D.}\ \bibnamefont
  {Kvon}}, \bibinfo {author} {\bibfnamefont {E.~B.}\ \bibnamefont
  {Olshanetsky}}, \bibinfo {author} {\bibfnamefont {D.~A.}\ \bibnamefont
  {Kozlov}}, \bibinfo {author} {\bibfnamefont {N.~N.}\ \bibnamefont
  {Mikhailov}}, \ and\ \bibinfo {author} {\bibfnamefont {S.~A.}\ \bibnamefont
  {Dvoretskii}},\ }\href {\doibase 10.1134/S0021364008090117} {\bibfield
  {journal} {\bibinfo  {journal} {JETP Lett.}\ }\textbf {\bibinfo {volume}
  {87}},\ \bibinfo {pages} {502} (\bibinfo {year} {2008})}\BibitemShut
  {NoStop}%
\bibitem [{\citenamefont {Kvon}\ \emph {et~al.}(2011)\citenamefont {Kvon},
  \citenamefont {Olshanetsky}, \citenamefont {Novik}, \citenamefont {Kozlov},
  \citenamefont {Mikhailov}, \citenamefont {Parm},\ and\ \citenamefont
  {Dvoretsky}}]{q49}%
  \BibitemOpen
  \bibfield  {author} {\bibinfo {author} {\bibfnamefont {Z.~D.}\ \bibnamefont
  {Kvon}}, \bibinfo {author} {\bibfnamefont {E.~B.}\ \bibnamefont
  {Olshanetsky}}, \bibinfo {author} {\bibfnamefont {E.~G.}\ \bibnamefont
  {Novik}}, \bibinfo {author} {\bibfnamefont {D.~A.}\ \bibnamefont {Kozlov}},
  \bibinfo {author} {\bibfnamefont {N.~N.}\ \bibnamefont {Mikhailov}}, \bibinfo
  {author} {\bibfnamefont {I.~O.}\ \bibnamefont {Parm}}, \ and\ \bibinfo
  {author} {\bibfnamefont {S.~A.}\ \bibnamefont {Dvoretsky}},\ }\href {\doibase
  10.1103/PhysRevB.83.193304} {\bibfield  {journal} {\bibinfo  {journal} {Phys.
  Rev. B}\ }\textbf {\bibinfo {volume} {83}},\ \bibinfo {pages} {193304}
  (\bibinfo {year} {2011})}\BibitemShut {NoStop}%
\bibitem [{\citenamefont {Hwang}\ and\ \citenamefont {Das~Sarma}(2008)}]{q51}%
  \BibitemOpen
  \bibfield  {author} {\bibinfo {author} {\bibfnamefont {E.~H.}\ \bibnamefont
  {Hwang}}\ and\ \bibinfo {author} {\bibfnamefont {S.}~\bibnamefont
  {Das~Sarma}},\ }\href {\doibase 10.1103/PhysRevB.77.195412} {\bibfield
  {journal} {\bibinfo  {journal} {Phys. Rev. B}\ }\textbf {\bibinfo {volume}
  {77}},\ \bibinfo {pages} {195412} (\bibinfo {year} {2008})}\BibitemShut
  {NoStop}%
\bibitem [{\citenamefont {Tkachov}\ \emph {et~al.}(2011)\citenamefont
  {Tkachov}, \citenamefont {Thienel}, \citenamefont {Pinneker}, \citenamefont
  {B\"uttner}, \citenamefont {Br\"une}, \citenamefont {Buhmann}, \citenamefont
  {Molenkamp},\ and\ \citenamefont {Hankiewicz}}]{q50}%
  \BibitemOpen
  \bibfield  {author} {\bibinfo {author} {\bibfnamefont {G.}~\bibnamefont
  {Tkachov}}, \bibinfo {author} {\bibfnamefont {C.}~\bibnamefont {Thienel}},
  \bibinfo {author} {\bibfnamefont {V.}~\bibnamefont {Pinneker}}, \bibinfo
  {author} {\bibfnamefont {B.}~\bibnamefont {B\"uttner}}, \bibinfo {author}
  {\bibfnamefont {C.}~\bibnamefont {Br\"une}}, \bibinfo {author} {\bibfnamefont
  {H.}~\bibnamefont {Buhmann}}, \bibinfo {author} {\bibfnamefont {L.~W.}\
  \bibnamefont {Molenkamp}}, \ and\ \bibinfo {author} {\bibfnamefont {E.~M.}\
  \bibnamefont {Hankiewicz}},\ }\href {\doibase 10.1103/PhysRevLett.106.076802}
  {\bibfield  {journal} {\bibinfo  {journal} {Phys. Rev. Lett.}\ }\textbf
  {\bibinfo {volume} {106}},\ \bibinfo {pages} {076802} (\bibinfo {year}
  {2011})}\BibitemShut {NoStop}%
\bibitem [{\citenamefont {Krishtopenko}\ \emph
  {et~al.}(2016{\natexlab{b}})\citenamefont {Krishtopenko}, \citenamefont
  {Knap},\ and\ \citenamefont {Teppe}}]{SMadd1}%
  \BibitemOpen
  \bibfield  {author} {\bibinfo {author} {\bibfnamefont {S.~S.}\ \bibnamefont
  {Krishtopenko}}, \bibinfo {author} {\bibfnamefont {W.}~\bibnamefont {Knap}},
  \ and\ \bibinfo {author} {\bibfnamefont {F.}~\bibnamefont {Teppe}},\ }\href
  {\doibase 10.1038/srep30755} {\bibfield  {journal} {\bibinfo  {journal} {Sci.
  Rep.}\ }\textbf {\bibinfo {volume} {6}},\ \bibinfo {pages} {30755} (\bibinfo
  {year} {2016}{\natexlab{b}})}\BibitemShut {NoStop}%
\end{thebibliography}

\begin{thebibliography}{7}%
\makeatletter
\providecommand \@ifxundefined [1]{%
 \@ifx{#1\undefined}
}%
\providecommand \@ifnum [1]{%
 \ifnum #1\expandafter \@firstoftwo
 \else \expandafter \@secondoftwo
 \fi
}%
\providecommand \@ifx [1]{%
 \ifx #1\expandafter \@firstoftwo
 \else \expandafter \@secondoftwo
 \fi
}%
\providecommand \natexlab [1]{#1}%
\providecommand \enquote  [1]{``#1''}%
\providecommand \bibnamefont  [1]{#1}%
\providecommand \bibfnamefont [1]{#1}%
\providecommand \citenamefont [1]{#1}%
\providecommand \href@noop [0]{\@secondoftwo}%
\providecommand \href [0]{\begingroup \@sanitize@url \@href}%
\providecommand \@href[1]{\@@startlink{#1}\@@href}%
\providecommand \@@href[1]{\endgroup#1\@@endlink}%
\providecommand \@sanitize@url [0]{\catcode `\\12\catcode `\$12\catcode
  `\&12\catcode `\#12\catcode `\^12\catcode `\_12\catcode `\%12\relax}%
\providecommand \@@startlink[1]{}%
\providecommand \@@endlink[0]{}%
\providecommand \url  [0]{\begingroup\@sanitize@url \@url }%
\providecommand \@url [1]{\endgroup\@href {#1}{\urlprefix }}%
\providecommand \urlprefix  [0]{URL }%
\providecommand \Eprint [0]{\href }%
\providecommand \doibase [0]{http://dx.doi.org/}%
\providecommand \selectlanguage [0]{\@gobble}%
\providecommand \bibinfo  [0]{\@secondoftwo}%
\providecommand \bibfield  [0]{\@secondoftwo}%
\providecommand \translation [1]{[#1]}%
\providecommand \BibitemOpen [0]{}%
\providecommand \bibitemStop [0]{}%
\providecommand \bibitemNoStop [0]{.\EOS\space}%
\providecommand \EOS [0]{\spacefactor3000\relax}%
\providecommand \BibitemShut  [1]{\csname bibitem#1\endcsname}%
\let\auto@bib@innerbib\@empty
\bibitem [{\citenamefont {Bernevig}\ \emph {et~al.}(2006)\citenamefont
  {Bernevig}, \citenamefont {Hughes},\ and\ \citenamefont {Zhang}}]{sm1}%
  \BibitemOpen
  \bibfield  {author} {\bibinfo {author} {\bibfnamefont {B.~A.}\ \bibnamefont
  {Bernevig}}, \bibinfo {author} {\bibfnamefont {T.~L.}\ \bibnamefont
  {Hughes}}, \ and\ \bibinfo {author} {\bibfnamefont {S.-C.}\ \bibnamefont
  {Zhang}},\ }\href {\doibase 10.1126/science.1133734} {\bibfield  {journal}
  {\bibinfo  {journal} {Science}\ }\textbf {\bibinfo {volume} {314}},\ \bibinfo
  {pages} {1757} (\bibinfo {year} {2006})}\BibitemShut {NoStop}%
\bibitem [{\citenamefont {K\"{o}nig}\ \emph {et~al.}(2008)\citenamefont
  {K\"{o}nig}, \citenamefont {Buhmann}, \citenamefont {W.~Molenkamp},
  \citenamefont {Hughes}, \citenamefont {Liu}, \citenamefont {Qi},\ and\
  \citenamefont {Zhang}}]{sm2}%
  \BibitemOpen
  \bibfield  {author} {\bibinfo {author} {\bibfnamefont {M.}~\bibnamefont
  {K\"{o}nig}}, \bibinfo {author} {\bibfnamefont {H.}~\bibnamefont {Buhmann}},
  \bibinfo {author} {\bibfnamefont {L.}~\bibnamefont {W.~Molenkamp}}, \bibinfo
  {author} {\bibfnamefont {T.}~\bibnamefont {Hughes}}, \bibinfo {author}
  {\bibfnamefont {C.-X.}\ \bibnamefont {Liu}}, \bibinfo {author} {\bibfnamefont
  {X.-L.}\ \bibnamefont {Qi}}, \ and\ \bibinfo {author} {\bibfnamefont {S.-C.}\
  \bibnamefont {Zhang}},\ }\href {\doibase 10.1143/JPSJ.77.031007} {\bibfield
  {journal} {\bibinfo  {journal} {J. Phys. Soc. Jpn.}\ }\textbf {\bibinfo
  {volume} {77}},\ \bibinfo {pages} {031007} (\bibinfo {year}
  {2008})}\BibitemShut {NoStop}%
\bibitem [{\citenamefont {Rothe}\ \emph {et~al.}(2010)\citenamefont {Rothe},
  \citenamefont {Reinthaler}, \citenamefont {Liu}, \citenamefont {Molenkamp},
  \citenamefont {Zhang},\ and\ \citenamefont {Hankiewicz}}]{sm3}%
  \BibitemOpen
  \bibfield  {author} {\bibinfo {author} {\bibfnamefont {D.~G.}\ \bibnamefont
  {Rothe}}, \bibinfo {author} {\bibfnamefont {R.~W.}\ \bibnamefont
  {Reinthaler}}, \bibinfo {author} {\bibfnamefont {C.-X.}\ \bibnamefont {Liu}},
  \bibinfo {author} {\bibfnamefont {L.~W.}\ \bibnamefont {Molenkamp}}, \bibinfo
  {author} {\bibfnamefont {S.-C.}\ \bibnamefont {Zhang}}, \ and\ \bibinfo
  {author} {\bibfnamefont {E.~M.}\ \bibnamefont {Hankiewicz}},\ }\href
  {\doibase 10.1088/1367-2630/12/6/065012} {\bibfield  {journal} {\bibinfo
  {journal} {New J. Phys.}\ }\textbf {\bibinfo {volume} {12}},\ \bibinfo
  {pages} {065012} (\bibinfo {year} {2010})}\BibitemShut {NoStop}%
\bibitem [{\citenamefont {Hwang}\ and\ \citenamefont {Das~Sarma}(2008)}]{sm4}%
  \BibitemOpen
  \bibfield  {author} {\bibinfo {author} {\bibfnamefont {E.~H.}\ \bibnamefont
  {Hwang}}\ and\ \bibinfo {author} {\bibfnamefont {S.}~\bibnamefont
  {Das~Sarma}},\ }\href {\doibase 10.1103/PhysRevB.77.195412} {\bibfield
  {journal} {\bibinfo  {journal} {Phys. Rev. B}\ }\textbf {\bibinfo {volume}
  {77}},\ \bibinfo {pages} {195412} (\bibinfo {year} {2008})}\BibitemShut
  {NoStop}%
\bibitem [{\citenamefont {Krishtopenko}\ and\ \citenamefont
  {Teppe}(2018)}]{sm5}%
  \BibitemOpen
  \bibfield  {author} {\bibinfo {author} {\bibfnamefont {S.~S.}\ \bibnamefont
  {Krishtopenko}}\ and\ \bibinfo {author} {\bibfnamefont {F.}~\bibnamefont
  {Teppe}},\ }\href {\doibase 10.1103/PhysRevB.97.165408} {\bibfield  {journal}
  {\bibinfo  {journal} {Phys. Rev. B}\ }\textbf {\bibinfo {volume} {97}},\
  \bibinfo {pages} {165408} (\bibinfo {year} {2018})}\BibitemShut {NoStop}%
\bibitem [{\citenamefont {Krishtopenko}\ \emph
  {et~al.}(2016{\natexlab{a}})\citenamefont {Krishtopenko}, \citenamefont
  {Yahniuk}, \citenamefont {But}, \citenamefont {Gavrilenko}, \citenamefont
  {Knap},\ and\ \citenamefont {Teppe}}]{sm6}%
  \BibitemOpen
  \bibfield  {author} {\bibinfo {author} {\bibfnamefont {S.~S.}\ \bibnamefont
  {Krishtopenko}}, \bibinfo {author} {\bibfnamefont {I.}~\bibnamefont
  {Yahniuk}}, \bibinfo {author} {\bibfnamefont {D.~B.}\ \bibnamefont {But}},
  \bibinfo {author} {\bibfnamefont {V.~I.}\ \bibnamefont {Gavrilenko}},
  \bibinfo {author} {\bibfnamefont {W.}~\bibnamefont {Knap}}, \ and\ \bibinfo
  {author} {\bibfnamefont {F.}~\bibnamefont {Teppe}},\ }\href {\doibase
  10.1103/PhysRevB.94.245402} {\bibfield  {journal} {\bibinfo  {journal} {Phys.
  Rev. B}\ }\textbf {\bibinfo {volume} {94}},\ \bibinfo {pages} {245402}
  (\bibinfo {year} {2016}{\natexlab{a}})}\BibitemShut {NoStop}%
\bibitem [{\citenamefont {Krishtopenko}\ \emph
  {et~al.}(2016{\natexlab{b}})\citenamefont {Krishtopenko}, \citenamefont
  {Knap},\ and\ \citenamefont {Teppe}}]{sm3b}%
  \BibitemOpen
  \bibfield  {author} {\bibinfo {author} {\bibfnamefont {S.~S.}\ \bibnamefont
  {Krishtopenko}}, \bibinfo {author} {\bibfnamefont {W.}~\bibnamefont {Knap}},
  \ and\ \bibinfo {author} {\bibfnamefont {F.}~\bibnamefont {Teppe}},\ }\href
  {\doibase 10.1038/srep30755} {\bibfield  {journal} {\bibinfo  {journal} {Sci.
  Rep.}\ }\textbf {\bibinfo {volume} {6}},\ \bibinfo {pages} {30755} (\bibinfo
  {year} {2016}{\natexlab{b}})}\BibitemShut {NoStop}%
\end{thebibliography}
%

\newpage
\clearpage
\setcounter{equation}{0}
\setcounter{figure}{0}
\setcounter{table}{0}
\renewcommand{\thefigure}{S\arabic{figure}}

\onecolumngrid
\section*{Supplemental Materials}
\maketitle
\onecolumngrid

\subsection{A. Self-consistent Born approximation within the two-band BHZ model.}
Electronic states in HgTe QWs in the vicinity of the $\Gamma$ point of the  Brillouin zone are qualitatively described by the BHZ Hamiltonian~\cite{sm1} for the lowest electron-like \emph{E}1 and top hole-like \emph{H}1 subbands.  Using the basis states $|E1,+\rangle$, $|H1,+\rangle$, $|E1,-\rangle$, $|H1,-\rangle$, the Hamiltonian for the \emph{E}1 and \emph{H}1 subbands is written as
\begin{equation}
\label{eq:sm1}
H_{2D}(\mathbf{k})=\begin{pmatrix}
H_{\mathrm{BHZ}}(\mathbf{k}) & 0 \\ 0 & H_{\mathrm{BHZ}}^{*}(-\mathbf{k})\end{pmatrix},
\end{equation}
where asterisk stands for complex conjugation, $\mathbf{k}=(k_x,k_y)$ is the momentum in the QW plane, and $H_{\mathrm{BHZ}}(\mathbf{k})=\epsilon_{k}\mathbf{I}_2+d_a(\mathbf{k})\sigma_a$ is the BHZ Hamiltonian~\cite{sm1}. Here, $\mathbf{I}_2$ is a 2$\times$2 unit matrix, $\sigma_a$ are the Pauli matrices, $\epsilon_{k}=C-D(k_x^2+k_y^2)$, $d_1(\mathbf{k})=Ak_x$, $d_2(\mathbf{k})=-Ak_y$, and  $d_3(k)=M-B(k_x^2+k_y^2)$. The structure parameters $C$, $M$, $A$, $B$, $D$ depend on $d$, strain, the barrier material, temperature and hydrostatic pressure. The mass parameter $M$ describes inversion between the \emph{E}1 and \emph{H}1 subbands. We note that $H_{2D}(\mathbf{k})$ has a block-diagonal form because the terms, which break inversion symmetry and axial symmetry around the growth direction, are neglected~\cite{sm2,sm3}. Further, we focus on the upper block only, while all the calculation for the lower block are performed in the similar manner.

Let us consider Green's function defined by
\begin{equation}
\label{eq:sm2}
\hat{G}(\mathbf{k},\varepsilon)=\langle\dfrac{1}{\varepsilon-\mathcal{H}}\rangle=
\left[\varepsilon-H_{\mathrm{BHZ}}(\mathbf{k})-\hat{\Sigma}(\mathbf{k},\varepsilon)\right]^{-1},
\end{equation}
with
\begin{equation}
\label{eq:sm3}
\mathcal{H}=H_{\mathrm{BHZ}}(\mathbf{k})+V_{imp}(\textbf{r}),
\end{equation}
where $\langle...\rangle$ denotes average over all disorder configurations, $\hat{\Sigma}(\mathbf{k},\varepsilon)$ is the self-energy matrix, and $V_{imp}(\textbf{r})$ is the disorder potential of the scatterers
\begin{equation}
\label{eq:sm4}
V_{imp}(\textbf{r})=\sum_{j}v(\textbf{r}-\textbf{R}_j).
\end{equation}
Here, $v(\textbf{r})$ is the potential of the scatter with the coordinate $R_j$. We consider the scatterers with isotropic potential
\begin{equation}
\label{eq:sm5}
v(\textbf{r})=\int\dfrac{d^2\textbf{q}}{(2\pi)^2}\tilde{v}(\textbf{q})e^{i\textbf{q}\cdot\textbf{r}},
\end{equation}
where $\tilde{v}(\textbf{q})=\tilde{v}(q)$ with $|\textbf{q}|=q$.

Due to full rotational symmetry of $H_{\mathrm{BHZ}}(\mathbf{k})$, its wave-function can be presented in the form:
\begin{equation}
\label{eq:sm6}
\Psi_{\mathrm{BHZ}}(\mathbf{k})=U(\theta_{\mathbf{k}})^{-1}\Psi_{\mathrm{BHZ}}(k),
\end{equation}
where $k=|\textbf{k}|$, $k_x=k\cos\theta_{\mathbf{k}}$, $k_y=k\sin\theta_{\mathbf{k}}$, and
\begin{equation}
\label{eq:sm7}
U(\theta)=\begin{pmatrix}
1 & 0 \\ 0 & e^{i\theta}\end{pmatrix}.
\end{equation}
Therefore, the Green's function in Eq.~(\ref{eq:sm2}) can be presented in the form
\begin{equation}
\label{eq:sm8}
\hat{G}(\mathbf{k},\varepsilon)=U(\theta_{\mathbf{k}})\hat{G}(k,\varepsilon)U(\theta_{\mathbf{k}})^{-1},
\end{equation}
with
\begin{equation}
\label{eq:sm9}
\hat{G}(k,\varepsilon)=\left[\varepsilon-\tilde{H}_{\mathrm{BHZ}}(k)-\hat{\Sigma}(k,\varepsilon)\right]^{-1},
\end{equation}
which depends only on $k$. This shows that $\hat{G}(\mathbf{k},\varepsilon)$ depends on the angle via the terms of $U(\theta_{\mathbf{k}})$. We note that $\tilde{H}_{\mathrm{BHZ}}(k)$ differs from $H_{\mathrm{BHZ}}(\mathbf{k})$ by
\begin{equation}
\label{eq:sm10}
\tilde{H}_{\mathrm{BHZ}}(k)=U(\theta_{\mathbf{k}})H_{\mathrm{BHZ}}(\mathbf{k})U(\theta_{\mathbf{k}})^{-1}.
\end{equation}

Within the SCBA, the self-energy matrix has a form:
\begin{equation}
\label{eq:sm11}
\hat{\Sigma}(\textbf{k},\varepsilon)=n_{i}\int\dfrac{d^2\textbf{k}^\prime}{(2\pi)^2}
\tilde{v}(\textbf{k}-\textbf{k}^\prime)\hat{G}(\mathbf{k}^\prime,\varepsilon)\tilde{v}(\textbf{k}^\prime-\textbf{k}),
\end{equation}
where $n_{i}$ is the concentration of impurities. By using Eq.~(\ref{eq:sm8}), we have
\begin{equation}
\label{eq:sm12}
\hat{\Sigma}(\textbf{k},\varepsilon)=n_{i}U(\theta_{\mathbf{k}})\int\dfrac{d^2\textbf{k}^\prime}{(2\pi)^2}
\tilde{v}(\textbf{k}-\textbf{k}^\prime)U(\theta_{\mathbf{k}^\prime}-\theta_{\mathbf{k}})\hat{G}(k^\prime,\varepsilon)U(\theta_{\mathbf{k}^\prime}-\theta_{\mathbf{k}})^{-1}\tilde{v}(\textbf{k}^\prime-\textbf{k})U(\theta_{\mathbf{k}})^{-1}.
\end{equation}
Thus, similar to Eq.~(\ref{eq:sm8}), the self-energy matrix can be written as
\begin{equation}
\label{eq:sm13}
\hat{\Sigma}(\mathbf{k},\varepsilon)=U(\theta_{\mathbf{k}})\hat{\Sigma}(k,\varepsilon)U(\theta_{\mathbf{k}})^{-1},
\end{equation}
where matrix $\hat{\Sigma}(k,\varepsilon)$ has a form
\begin{equation}
\label{eq:sm14}
\hat{\Sigma}(k,\varepsilon)=n_{i}\int\limits_0^{K_c}\dfrac{k^\prime dk^\prime}{2\pi}\begin{pmatrix}
V_0(k,k^\prime)^2G_{11}^\prime & V_{-1}(k,k^\prime)^2G_{12}^\prime \\ V_{+1}(k,k^\prime)^2G_{21}^\prime & V_0(k,k^\prime)^2G_{22}^\prime \end{pmatrix},
\end{equation}
where $G_{ij}^\prime\equiv G_{ij}(k^\prime,\varepsilon)$ are the component of the Green's function in Eq.~(\ref{eq:sm8}), and
$V_n(k,k^\prime)^2$ is written as
\begin{equation}
\label{eq:sm15}
V_n(k,k^\prime)^2=\int\limits_0^{2\pi}\dfrac{d\theta}{2\pi}|\tilde{v}(\textbf{k}-\textbf{k}^\prime)|^2
\cos n\theta.
\end{equation}
In Eq.~(\ref{eq:sm14}), we introduce a cut-off wave-vector $K_c=\pi/a_{\mathrm{0}}$ (where $a_{\mathrm{0}}$ is the lattice constant), which corresponds to the size of the first Brillouin zone. Once the self-energy is known, we can express the spectral function $A(k,\varepsilon)$ and the density-of-states $D(\varepsilon)$:
\begin{eqnarray}
\label{eq:sm16}
A(k,\varepsilon)=-\dfrac{1}{\pi}\textrm{Im}\left\{\textrm{Tr}\left(\hat{G}(k,\varepsilon+i0)\right)\right\},\notag\\
D(\varepsilon)=g_S\int\limits_0^{K_c}\dfrac{k dk}{2\pi}A(k,\varepsilon),~~~~~~~~~~
\end{eqnarray}
where the factor $g_S=2$ takes into account the contribution from the lower block in Eq.~(\ref{eq:sm1}).

In the case of the short-range impurities, $\tilde{v}(q)=u_0$, and the self-energy matrix is independent of $k$ and has the form $\hat{\Sigma}(\varepsilon)=\Sigma_0(\varepsilon)\mathbf{I}_2+\Sigma_z(\varepsilon)\sigma_z$. As a results, the set in Eq.~(\ref{eq:sm14}) is written as
\begin{eqnarray}
\label{eq:sm17}
\Sigma_0(\varepsilon)=\dfrac{W^2}{4\pi}\int\limits_0^{K_c^2}
\dfrac{X(\varepsilon)+Dx}
{\Lambda(x,\varepsilon)}dx,~~~~
\Sigma_z(\varepsilon)=\dfrac{W^2}{4\pi}\int\limits_0^{K_c^2}
\dfrac{Y(\varepsilon)-Bx}
{\Lambda(x,\varepsilon)}dx,~~~~\notag\\
\Lambda(x,\varepsilon)=\left(D^2-B^2\right)x^2+\left(2BY(\varepsilon)+2DX(\varepsilon)-A^2\right)x+X(\varepsilon)^2-Y(\varepsilon)^2,~\notag\\
X(\varepsilon)=\varepsilon-C-\Sigma_0(\varepsilon),~~~~~Y(\varepsilon)=M+\Sigma_z(\varepsilon),~~~~~~~~~~~~~~~~~~
\end{eqnarray}
where the disorder strength $W$ is defined as $W^2=n_{i}u_0^2$. For the case of the short-range impurities, $A(k,\varepsilon)$ and $D(\varepsilon)$ are written as
\begin{eqnarray}
\label{eq:sm17a}
A(k,\varepsilon)=-\dfrac{1}{\pi}\textrm{Im}\left\{
\dfrac{2\left(\varepsilon-\epsilon_{k}-\Sigma_0(\varepsilon)\right)}{\left(\varepsilon-\epsilon_{k}-\Sigma_0(\varepsilon)\right)^2
-\left(d_3(k)+\Sigma_z(\varepsilon)\right)^2-A^2k^2}
\right\},\notag\\
D(\varepsilon)=-\dfrac{2g_S}{W^2\pi}\textrm{Im}\left\{\Sigma_0(\varepsilon)\right\},~~~~~~~~~~~~~~~~~~~~~~~~~~
\end{eqnarray}

Let us first consider the case, when $|B|=|D|$, which makes $\Lambda(x,\varepsilon)$ a linear function of $x$. Under these conditions, the integrals for $\hat{\Sigma}(\varepsilon)$ and $\Sigma_z(\varepsilon)$ have a form:
\begin{equation}
\label{eq:sm18}
\int\dfrac{ax+b}{cx+d}dx=\dfrac{a}{c}x+\dfrac{bc-ad}{c^2}\ln\left(cx+d\right),
\end{equation}
which is valid even for the complex values of $a$, $b$, $c$ and $d$. Note that the linear approximation for the Dirac fermions in graphene, i.e. $B=D=0$ corresponds to $a=0$.

If $|B|\neq|D|$, $\Lambda(x,\varepsilon)$ can be always presented in the form $\Lambda(x,\varepsilon)=c(x-x_1)(x-x_2)$, where $x_1$ and $x_2$ are the roots of the square polinom. We note the complex values of $c$, $x_1$ and $x_2$ in general case. As a result,
the calculation of $\hat{\Sigma}(\varepsilon)$ and $\Sigma_z(\varepsilon)$ is reduced to the calculation of the integrals:
\begin{equation}
\label{eq:sm19}
\int\dfrac{ax+b}{c(x-x_1)(x-x_2)}dx=\dfrac{ax_1+b}{c(x_1-x_2)}\ln\left(x-x_1\right)
+\dfrac{ax_2+b}{c(x_2-x_1)}\ln\left(x-x_2\right).
\end{equation}
By using Eqs.~(\ref{eq:sm18}) and (\ref{eq:sm19}), Eq.~(\ref{eq:sm17}) transforms into the set of algebraic equations numerically solved by iteration procedure. To calculate $\hat{\Sigma}(\varepsilon)$ and $\Sigma_z(\varepsilon)$ at the $n$th iteration step, we use their values for the right parts determined at the $(n-1)$th iteration. The maximum number of iterations $n_{\mathrm{max}}$ in the numerical calculations was 1000. For the zeroth iteration step, $X^{(0)}(\varepsilon)=\varepsilon+i0$ and $Y^{(0)}(\varepsilon)=M$.

\begin{figure}
\includegraphics [width=0.70\columnwidth, keepaspectratio] {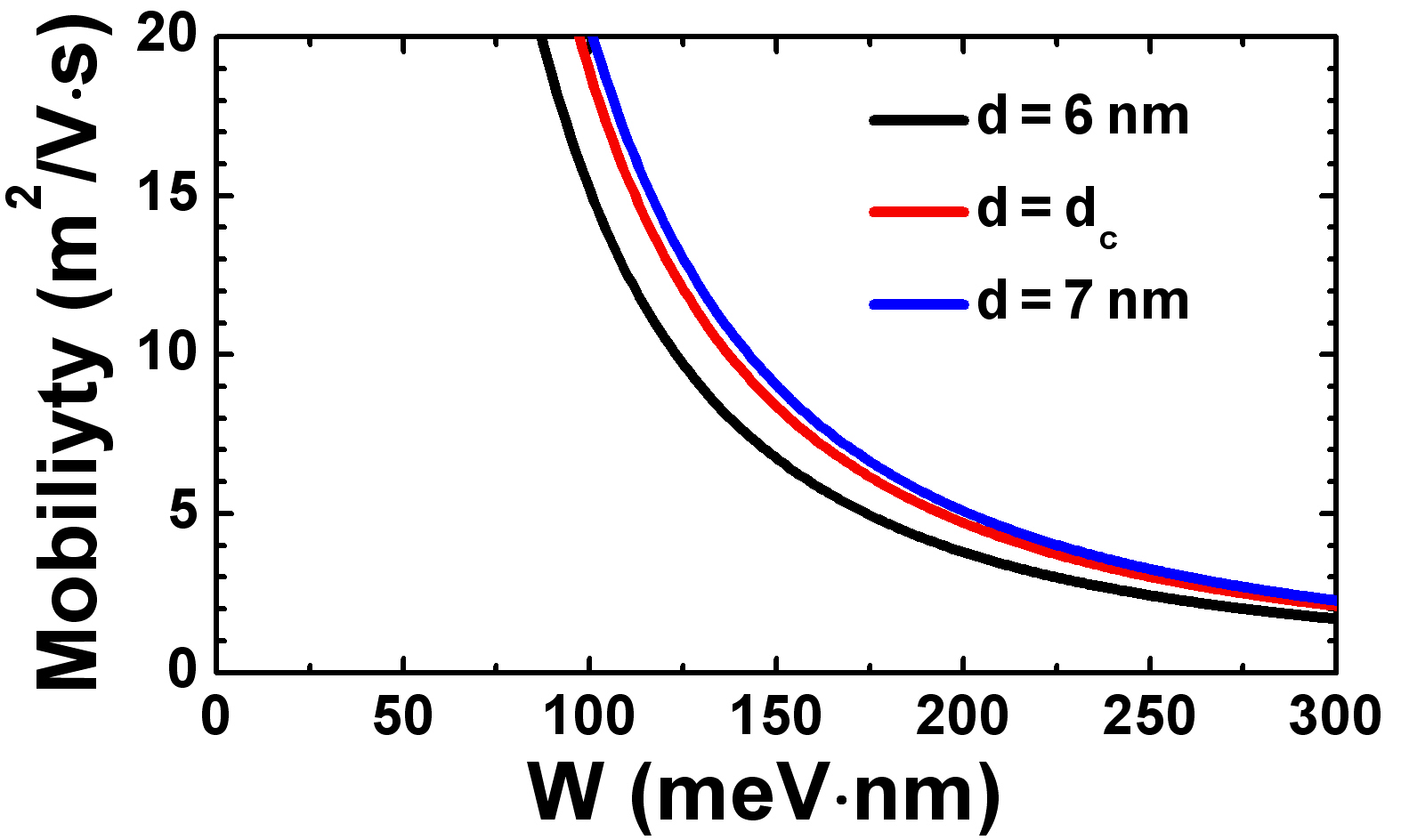} 
\caption{\label{Fig:SM1} Electron mobility $\mu_{W}$ as a function of the disorder strength $W$ calculated in the two-band BHZ model for HgTe QW at different QW width: $d=6$~nm, $d=d_c$ and $d=7$~nm for the electron concentration $n_S=10^{11}$~cm$^{-2}$ (see the main text).}
\end{figure}

As mentioned in the main text, the disorder strength $W$ is connected with the electron mobility values, which can be calculated for the certain types of the disorder. Particularly, the electron mobility $\mu_{W}$ at $T=0$~K caused by the short-range impurities can be evaluated in the relaxation time approximation~\cite{sm4}:
\begin{eqnarray}
\label{eq:sm20}
\dfrac{\hbar}{\tau_{tr}^{(W)}}=2\pi n_{i}\sum\limits_{\lambda^\prime}\int\dfrac{d^2\textbf{k}^\prime}{(2\pi)^2}
\left|\tilde{v}(\textbf{k}-\textbf{k}^\prime)\right|^2F_{\lambda\lambda^\prime}(\textbf{k},\textbf{k}^\prime)
\left(1-\cos(\theta_{\mathbf{k}^\prime}-\theta_{\mathbf{k}})\right)\delta\left(E_F-E_{\lambda^\prime,k^\prime}\right),~~~~
\end{eqnarray}
with
\begin{eqnarray}
\label{eq:sm21}
F_{\lambda\lambda^\prime}(\textbf{k},\textbf{k}^\prime)=\dfrac{1}{2}\left(1+\lambda\lambda^\prime\hat{d}_{\textbf{k}}\hat{d}_{\textbf{k}^\prime}\right), \notag\\
\hat{d}_{\textbf{k}}\hat{d}_{\textbf{k}^\prime}=\dfrac{A^2k^2\cos(\theta_{\mathbf{k}^\prime}-\theta_{\mathbf{k}})+d^2_3(k)}{A^2k^2+d^2_3(k)},\notag\\
E_{\lambda,k}=\epsilon_{k}+\lambda\sqrt{A^2k^2+d^2_3(k)},
\end{eqnarray}
where $\lambda$,$\lambda^\prime=\pm 1$ denote the indices for conduction ($+1$) and valence ($-1$) band. Assuming that Fermi level lies in the conduction band ($\lambda=+1$) and defining Fermi vector as $k_F=\sqrt{2\pi n_s}$ ($E_F=E_{\lambda=+1,k_F}$), the straight calculations results in\begin{eqnarray}
\label{eq:sm22}
\dfrac{\hbar}{\tau_{tr}^{(W)}}=\dfrac{W^{2}}{\hbar^2/m_c(k_F)}\dfrac{A^2k_F^2/4+d_3^2(k_F)}{A^2k_F^2+d_3^2(k_F)},
\end{eqnarray}
where we have introduced a cyclotron mass $m_c(k_F)$ at the Fermi level:
\begin{eqnarray}
\label{eq:sm23}
\dfrac{1}{m_c}=\dfrac{1}{\hbar^2k_F}\dfrac{\partial E_{\lambda=+1,k}}{\partial k}\bigg|_{k=k_F}.
\end{eqnarray}
Then, if $\tau_{tr}^{(W)}$ is known, the mobility is calculated as $\mu_{W}=e\tau_{tr}^{(W)}/m_c(k_F)$. Fig.~\ref{Fig:SM1} shows that for the 6~nm HgTe QW (see the main text) with $n_S=10^{11}$~cm$^{-2}$, $W<200$ corresponds to $\mu_{W}>4$~m$^2$/V$\cdot$s.

Note that HgTe QW may have other imperfections, which differ from the short-range impurities, also resulting to finite electron mobility $\mu_0$. In this case, the total mobility $\mu_{tot}$ including both contributions is calculated as
\begin{eqnarray}
\label{eq:sm24}
\mu_{tot}=\dfrac{\mu_{W}\mu_0}{\mu_{W}+\mu_0}.
\end{eqnarray}

\subsection{B. Self-consistent Born approximation within the four-band 2D Hamiltonian.}
The four-band 2D Hamiltonian $H_{2D}(\mathbf{k})$ for extended description of HgTe QWs~\cite{sm5} including the second electron-like \emph{E}2 and hole-like \emph{H}2 subbands in the basis $|$\emph{E}1,+$\rangle$, $|$\emph{H}1,+$\rangle$, $|$\emph{H}2,-$\rangle$, $|$\emph{E}2,-$\rangle$, $|$\emph{E}1,-$\rangle$, $|$\emph{H}1,-$\rangle$, $|$\emph{H}2,+$\rangle$, $|$\emph{E}2,-$\rangle$ has the form:
\begin{equation}
\label{eq:Bsm1}
H_{2D}(\mathbf{k})=\begin{pmatrix}
H_{4\times4}(\mathbf{k}) & 0 \\ 0 & H_{4\times4}^{*}(-\mathbf{k})\end{pmatrix}
\end{equation}
with the blocks $H_{4\times4}(\mathbf{k})$ and $H_{4\times4}^{*}(-\mathbf{k})$ defined as
\begin{equation}
\label{eq:Bsm2}
H_{4\times4}(\mathbf{k})=\begin{pmatrix}
\epsilon_{\mathbf{k}}+d_3(\mathbf{k}) & -Ak_{+} & R_{1}k_{-}^2 & S_{0}k_{-}\\
-Ak_{-} & \epsilon_{\mathbf{k}}-d_3(\mathbf{k}) & 0 & R_{2}k_{-}^2\\
R_{1}k_{+}^2 & 0 & \epsilon_{H2}(\mathbf{k})  & A_{2}k_{+}\\
S_{0}k_{+} & R_{2}k_{+}^2 & A_{2}k_{-} & \epsilon_{E2}(\mathbf{k}) \end{pmatrix},
\end{equation}
where $\epsilon_{E2}(\mathbf{k})=C+M+\Delta_{E1E2}+B_{E2}(k_x^2+k_y^2)$, $\epsilon_{H2}(\mathbf{k})=C-M-\Delta_{H1H2}+B_{H2}(k_x^2+k_y^2)$, $\Delta_{E1E2}$ and $\Delta_{H1H2}$ are the gaps between the \emph{E}1 and \emph{E}2 subbands and the \emph{H}1 and \emph{H}2 subbands, respectively~\cite{sm5}. Other parameters are the same as those for the BHZ Hamiltonian. As it is for the two-band BHZ model, we have also neglected the terms breaking inversion symmetry and axial symmetry around the growth direction~\cite{sm2,sm3}. This results in the block-diagonal form of $H_{2D}(\mathbf{k})$, each of them can be considered independently.

Due to full rotational symmetry of $H_{4\times4}(\mathbf{k})$, its wave-function can be presented in the form:
\begin{equation}
\label{eq:Bsm3}
\Psi_{4\times4}(\mathbf{k})=U(\theta_{\mathbf{k}})^{-1}\Psi_{4\times4}(k),
\end{equation}
where
\begin{equation}
\label{eq:Bsm4}
U(\theta)=\begin{pmatrix}
1 & 0 & 0 & 0 \\
0 & e^{i\theta} & 0 & 0 \\
0 & 0 & e^{-2i\theta} & 0 \\
0 & 0 & 0 & e^{-i\theta} \end{pmatrix}.
\end{equation}
Therefore, the averaged Green's function
\begin{equation*}
\hat{G}(\mathbf{k},\varepsilon)=\left[\varepsilon-H_{4\times4}(\mathbf{k})-\hat{\Sigma}(\mathbf{k},\varepsilon)\right]^{-1},
\end{equation*}
can be presented in the form $\hat{G}(\mathbf{k},\varepsilon)=U(\theta_{\mathbf{k}})\hat{G}(k,\varepsilon)U(\theta_{\mathbf{k}})^{-1}$,
with $\hat{G}(k,\varepsilon)$, which depends only on $k$. This shows that $\hat{G}(\mathbf{k},\varepsilon)$ depends on the angle via the terms of $U(\theta_{\mathbf{k}})$.

Thus, similar to Eq.~(\ref{eq:sm13}), the self-energy matrix for the $H_{4\times4}(\mathbf{k})$ Hamiltonian is be written as
\begin{equation}
\label{eq:Bsm5}
\hat{\Sigma}(\mathbf{k},\varepsilon)=U(\theta_{\mathbf{k}})\hat{\Sigma}(k,\varepsilon)U(\theta_{\mathbf{k}})^{-1},
\end{equation}
where matrix $\hat{\Sigma}(k,\varepsilon)$ has a form
\begin{equation}
\label{eq:Bsm6}
\hat{\Sigma}(k,\varepsilon)=n_{i}\int\limits_0^{K_c}\dfrac{k^\prime dk^\prime}{2\pi}\begin{pmatrix}
V_0^2G_{11}^\prime & V_{-1}^2G_{12}^\prime & V_{+2}^2G_{13}^\prime & V_{+1}^2G_{14}^\prime \\[4pt]
V_{+1}^2G_{21}^\prime & V_{0}^2G_{22}^\prime & V_{+3}^2G_{23}^\prime & V_{+2}^2G_{24}^\prime \\[4pt]
V_{-2}^2G_{31}^\prime & V_{-3}^2G_{32}^\prime & V_{0}^2G_{33}^\prime & V_{-1}^2G_{34}^\prime \\[4pt]
V_{-1}^2G_{41}^\prime & V_{-2}^2G_{42}^\prime & V_{+1}^2G_{43}^\prime & V_{0}^2G_{44}^\prime
\end{pmatrix},
\end{equation}
where $V_n^2\equiv V_n(k,k^\prime)^2$ is given by Eq.~(\ref{eq:sm15}) and $G_{ij}^\prime\equiv G_{ij}(k^\prime,\varepsilon)$ are the component of the averaged Green's function
\begin{equation}
\label{eq:Bsm7}
\hat{G}(k,\varepsilon)=\left[\varepsilon-\tilde{H}_{4\times4}(k)-\hat{\Sigma}(k,\varepsilon)\right]^{-1}.
\end{equation}
Here $H_{4\times4}(k)$ differs from $H_{4\times4}(\mathbf{k})$ by $H_{4\times4}(k)=U(\theta_{\mathbf{k}})H_{4\times4}(\mathbf{k})U(\theta_{\mathbf{k}})^{-1}$.

In the case of the short-range impurities, $V_n(k,k^\prime)^2=u_0^2\delta_{n,0}$, and the self-energy matrix in Eq.~(\ref{eq:Bsm6}) is independent of $k$ and has the diagonal form
\begin{equation}
\label{eq:Bsm8}
\hat{\Sigma}(\varepsilon)=\begin{pmatrix}
\Sigma_{E1}(\varepsilon) & 0 & 0 & 0 \\[4pt]
0 & \Sigma_{H1}(\varepsilon) & 0 & 0 \\[4pt]
0 & 0 & \Sigma_{H2}(\varepsilon) & 0 \\[4pt]
0 & 0 & 0 & \Sigma_{E2}(\varepsilon)
\end{pmatrix}=
\dfrac{W^2}{4\pi}\int\limits_0^{K_c^2}dx
\begin{pmatrix}
G_{11}\left(\sqrt{x},\varepsilon\right) & 0 & 0 & 0 \\[4pt]
0 & G_{22}\left(\sqrt{x},\varepsilon\right) & 0 & 0 \\[4pt]
0 & 0 & G_{33}\left(\sqrt{x},\varepsilon\right) & 0 \\[4pt]
0 & 0 & 0 & G_{44}\left(\sqrt{x},\varepsilon\right)
\end{pmatrix},
\end{equation}
where the disorder strength is defined as $W^2=n_{i}u_0^2$ (cf. Eq.~(\ref{eq:sm17})). The given form of the self-energy and its independence of $k$ allows for an analytical calculation of the integrals in Eq.~(\ref{eq:Bsm8}).

First, we note the diagonal form of the matrix $\left[\varepsilon-\tilde{H}_{4\times4}(\sqrt{x})-\hat{\Sigma}(\varepsilon)\right]$ in Eq.~(\ref{eq:Bsm7}), whose determinant is the four-degree polynomial with respect to $x$:
\begin{equation}
\label{eq:Bsm9}
\det\left(\varepsilon-\tilde{H}_{4\times4}(\sqrt{x})-\hat{\Sigma}(\varepsilon)\right)=A_4(\varepsilon)x^4+A_3(\varepsilon)x^3+A_2(\varepsilon)x^2+A_1(\varepsilon)x+A_0(\varepsilon).
\end{equation}
Explicit forms for $A_4(\varepsilon)$, $A_3(\varepsilon)$, $A_2(\varepsilon)$, $A_1(\varepsilon)$ and $A_0(\varepsilon)$ are found by straightforward calculation of $4\times4$ symmetric matrix determinant:
\begin{multline*}
\det\left(\hat{A}\right)=a_{12}^2a_{34}^2-a_{33}a_{44}a_{12}^2+2a_{44}a_{12}a_{13}a_{23}-2a_{12}a_{13}a_{24}a_{34}-2a_{12}a_{14}a_{23}a_{34}+2a_{33}a_{12}a_{14}a_{24}-a_{22}a_{33}a_{14}^2~~~~~~~~\\
+a_{13}^2a_{24}^2-a_{22}a_{44}a_{13}^2-2a_{13}a_{14}a_{23}a_{24}+2a_{22}a_{13}a_{14}a_{34}+a_{14}^2a_{23}^2
-a_{11}a_{44}a_{23}^2+2a_{11}a_{23}a_{24}a_{34}-a_{11}a_{33}a_{24}^2~\\
-a_{11}a_{22}a_{34}^2+a_{11}a_{22}a_{33}a_{44}.~~~~
\end{multline*}
Second, the diagonal components of the Green's function $G_{ii}\left(\sqrt{x},\varepsilon\right)$ ($i=1...4$) in Eq.~(\ref{eq:Bsm8}) are presented as
\begin{equation}
\label{eq:Bsm10}
G_{ii}\left(\sqrt{x},\varepsilon\right)=\dfrac{a_{i}(\varepsilon)x^3+b_{i}(\varepsilon)x^2+c_{i}(\varepsilon)x+d_{i}(\varepsilon)}{A_4(\varepsilon)x^4+A_3(\varepsilon)x^3+A_2(\varepsilon)x^2+A_1(\varepsilon)x+A_0(\varepsilon)}.
\end{equation}
The latter can be verified by the direct calculation of the inverse matrix $\left[\varepsilon-\tilde{H}_{4\times4}(\sqrt{x})-\hat{\Sigma}(\varepsilon)\right]^{-1}$. As the self-energy matrix has imaginary part, all the coefficients in Eq.~(\ref{eq:Bsm10}) are complex as well. Further, we do not mark their dependence on $\varepsilon$ and omit index $i$.

In order to calculate the integrals in Eq.~(\ref{eq:Bsm8}), we have numerically found the roots $x_1$, $x_2$, $x_3$, $x_4$ of the polynomial needed for the following expansion:
\begin{equation}
\label{eq:Bsm11}
A_4x^4+A_3x^3+A_2x^2+A_1x+A_0=A_4(x-x_1)(x-x_2)(x-x_3)(x-x_4).
\end{equation}
Although the values of $x_1$, $x_2$, $x_3$, $x_4$ can be found analytically by means of \emph{Ferrari's method}, the numerical procedures also allow for the calculations with any needed degree of accuracy. Once the roots are known, the integrals are calculated as
\begin{multline}
\label{eq:Bsm12}
\int\dfrac{ax^3+bx^2+cx+d}{A_4(x-x_1)(x-x_2)(x-x_3)(x-x_4)}dx=\dfrac{ax_1^3+bx_1^2+cx_1+d}{A_4(x_1-x_2)(x_1-x_3)(x_1-x_4)}\ln\left(x-x_1\right)~~~~~~~~~~~~~~~~~~~~~~~~~\\
+\dfrac{ax_2^3+bx_2^2+cx_2+d}{A_4(x_2-x_1)(x_2-x_3)(x_2-x_4)}\ln\left(x-x_2\right)
+\dfrac{ax_3^3+bx_3^2+cx_3+d}{A_4(x_3-x_1)(x_3-x_2)(x_3-x_4)}\ln\left(x-x_3\right)~~~~~~\\
+\dfrac{ax_4^3+bx_4^2+cx_4+d}{A_4(x_4-x_1)(x_4-x_2)(x_4-x_3)}\ln\left(x-x_4\right).
\end{multline}
Eqs.~(\ref{eq:Bsm11}) and (\ref{eq:Bsm12}) allow for transformation of Eq.~(\ref{eq:Bsm8}) into the set of algebraic equations numerically solved by iteration procedure, as described in the Section~A.

After the Green's function $\hat{G}(\varepsilon)$ and self-energy matrix $\hat{\Sigma}(\varepsilon)$ are known, the spectral function $A(k,\varepsilon)$ and density-of-states $D(\varepsilon)$ are calculated as
\begin{eqnarray}
\label{eq:Bsm12}
A(k,\varepsilon)=-\dfrac{1}{\pi}\textrm{Im}\left\{G_{11}\left(k,\varepsilon\right)
+G_{22}\left(k,\varepsilon\right)+G_{33}\left(k,\varepsilon\right)
+G_{44}\left(k,\varepsilon\right)\right\},\notag\\
D(\varepsilon)=-\dfrac{g_S}{W^2\pi}\textrm{Im}\left\{\Sigma_{E1}(\varepsilon)
+\Sigma_{H1}(\varepsilon)+\Sigma_{H2}(\varepsilon)
+\Sigma_{E2}(\varepsilon)\right\}.~~~~~~~
\end{eqnarray}
The latter is valid only for the case of short-range impurities.

\subsection{C. Parameters for the effective 2D models}

By using the 8-band Kane Hamiltonian, accounting interaction between the $\Gamma_6$, $\Gamma_8$ and $\Gamma_7$ bands in zinc-blend materials~\cite{sm6} and by applying the procedure, described in~Ref.~\cite{sm3b}, one can calculate parameters for the effective 2D models. Parameters for $H_{4\times4}(\mathbf{k})$ are given in Table~\ref{tab:1}. To obtain parameters for $H_{\mathrm{BHZ}}(\mathbf{k})$ Hamiltonian from those for $H_{4\times4}(\mathbf{k})$, one should renormalize $B$ and $D$ as follows:
\begin{equation}
\label{eq:SM6}
B^{(\mathrm{BHZ})}=B^{(4\times4)}-\dfrac{S_0^2}{2\Delta_{E1E2}},~~~~~~~D^{(\mathrm{BHZ})}=D^{(4\times4)}-\dfrac{S_0^2}{2\Delta_{E1E2}}.
\end{equation}

\begin{table*}\caption{\label{tab:1} Structure parameters for $H_{4\times4}(\mathbf{k})$.}
\begin{ruledtabular}
\begin{tabular}{cccccccc}
HgTe QW width & $C$~(meV) & $M$~(meV) & $B$~(meV$\cdot$nm$^2$) & $D$~(meV$\cdot$nm$^2$) & $A$~(meV$\cdot$nm) & $\Delta_{H1H2}$~(meV) & $\Delta_{E1E2}$~(meV) \\
\hline
6~nm                & -24.61 &  6.49  & -568  & -394  & 380 & 65.94 & 332.68 \\
$d_c\simeq 6.58$~nm & -28.16 &  0.00  & -673  & -499  & 370 & 57.16 & 312.06 \\
7~nm                & -30.64 & -4.53  & -768  & -593  & 363 & 51.14 & 297.57 \\
8~nm                & -35.19 & -12.58 & -994  & -820  & 347 & 40.70 & 269.47 \\
9~nm                & -38.59 & -18.51 & -1324 & -1149 & 330 & 33.23 & 246.7
\end{tabular}
\begin{tabular}{ccccccc}
HgTe QW width & $R_1$~(meV$\cdot$nm$^2$) & $R_{2}$~(meV$\cdot$nm$^2$) & $B_{H2}$~(meV$\cdot$nm$^2$) & $B_{E2}$~(meV$\cdot$nm$^2$) & $A_2$~(meV$\cdot$nm) & $S_0$~(meV$\cdot$nm) \\
\hline
6~nm                & -1067 & -42.7 & 919 & -21.4 & 450 & 35.4 \\
$d_c\simeq 6.58$~nm & -1017 & -43.1 & 776 & -26.6 & 441 & 41.0  \\
7~nm                & -1007 & -43.5 & 711 & -30.0 & 427 & 44.7 \\
8~nm                & -1050 & -44.4 & 619 & -35.0 & 381 & 51.9 \\
9~nm                & -1155 & -45.3 & 572 & -38.9 & 312 & 57.3
\end{tabular}
\end{ruledtabular}
\end{table*}

\end{document}